\documentclass[12pt, draftclsnofoot, onecolumn]{IEEEtran}
\usepackage[T1]{fontenc}
\usepackage[latin9]{inputenc}
\usepackage{array}
\usepackage{mathtools}
\usepackage{amsmath}
\usepackage{amssymb}
\usepackage{graphicx}
\usepackage{epsfig}
\usepackage{cite}
\usepackage{color}
\usepackage{pstool}
\usepackage{psfrag}

\usepackage{algpseudocode,algorithm}
\makeatletter
\newcommand{\algmargin}{\the\ALG@thistlm}
\DeclareMathOperator*{\argmax}{arg\,max} 
\DeclareMathOperator*{\argmin}{arg\,min}
\DeclareMathOperator*{\tr}{tr}
\makeatother
\newlength{\whilewidth}
\algdef{SE}[parWHILE]{parWhile}{EndparWhile}[1]
  {\parbox[t]{\dimexpr\linewidth-\algmargin}{%
     \hangindent\whilewidth\strut\algorithmicwhile\ #1\ \algorithmicdo\strut}}{\algorithmicend\ \algorithmicwhile}%
\algnewcommand{\parState}[1]{\State%
  \parbox[t]{\dimexpr\linewidth-\algmargin}{\strut #1\strut}}

\ifCLASSINFOpdf
\else
\fi
\usepackage[T1]{fontenc}
\usepackage[latin9]{inputenc}
\usepackage{amsmath}
\usepackage{amsthm}

\theoremstyle{plain}
\newtheorem{thm}{\protect\theoremname}
\theoremstyle{remark}
\newtheorem{rem}{\protect\remarkname}
\newtheorem{lem}{Lemma}
\newtheorem{prop}[thm]{\protect\propositionname}

\makeatother

\providecommand{\propositionname}{Proposition}
\providecommand{\remarkname}{Remark}
\providecommand{\theoremname}{Theorem}

\begin{document}
%
\title{Location-aware Beam Alignment for mmWave Communications}
%
%
%

\author{Igbafe~Orikumhi,~\IEEEmembership{Member,~IEEE,}
        Jeongwan~Kang,~\IEEEmembership{Student~Member,~IEEE,}
        Henk Wymeersch,~\IEEEmembership{Member,~IEEE,}
        and~Sunwoo~Kim,~\IEEEmembership{Senior~Member,~IEEE}
        \thanks{Igbafe~Orikumhi, Jeongwan~Kang, and~Sunwoo~Kim are with the Department of Electronics and Computer Engineering, Hanyang University, Seoul, South Korea, email: \{oigbafe2,rkdwjddhks77,remero\}@hayang.ac.kr. Henk Wymeersch is  with the Department of Signal and Systems, Chalmers University of Technology, Sweden, email: henkw@chalmers.se. This work was supported by Samsung Research Funding \& Incubation Center of Samsung Electronics under Project Number SRFC-IT-1601-09 and by the MSIT (Ministry of Science and ICT), Korea, under the ITRC (Information Technology Research Center) support program (IITP-2019-2017-0-01637) supervised by the IITP (Institute for Information \& Communications Technology Planning \& Evaluation). Part of this work was previously presented at the 2018 Annual Allerton Conference on Communication, Control, and Computing (Allerton) \cite{8635826}. 
}
}

\maketitle

\begin{abstract}
Beam alignment is required in millimeter wave communication to ensure
high data rate transmission. However, with narrow beamwidth in massive
MIMO, beam alignment could be computationally intensive due to the
large number of beam pairs to be measured. In this paper, we propose
an efficient beam alignment framework by exploiting the  location
information of the user equipment (UE) and potential reflecting points. The
proposed scheme allows the UE and the base station to perform a coordinated
beam search from a small set of beams within the error boundary of the
location information, the selected beams are then used to guide the
search of future beams. To further reduce the number of beams to be
searched, we propose an intelligent search scheme  within a small window of
beams to determine the direction of the actual beam. The proposed
beam alignment algorithm is verified on simulation with some location
uncertainty.
\end{abstract}


%
\IEEEpeerreviewmaketitle

\section{Introduction}
%
%
%
%

 

\IEEEPARstart{M}{illimeter-wave} (mmWave) spectrum has been proposed for the fifth generation (5G) communication networks
due to the large bandwidth available at this frequency band. Other
advantages such as beamforming and spatial multiplexing are leading
to an increased interest in the mmWave spectrum \cite{6515173}. However,
the key challenge is that the mmWave spectrum suffers from severe path-loss. On the other hand, the high-frequency spectrum allows the use
of a large antenna array with small form factor, which provides high
beamforming gains to compensate for the losses. As such, beamforming
at the base station (BS) and the user equipment (UE) have become an essential
part of the mmWave 5G networks \cite{8335545,8277251}.

Although as the number of array antenna increases, the array gain
also increases with reduced interference \cite{8385501}, a new challenge
of selecting the best beam pair at the transmitter and receiver exist
due to the narrow beamwidth \cite{7742901}, which could compromise
the transmission rate especially when the time taken for beam alignment
is large \cite{8370110}. Consequently, beam  alignment especially in a
mobile environment may become the bottleneck of the communication network.
More specifically, in the initial access phase, beams with narrow
beamwidth can complicate the initial cell search since the UE
and BS have to search over a large angular directional space
for   suitable path to establish   communication \cite{7914759,7947209}.


When a user enters a cell, the user establishes a physical
connection to the BS in the initial access phase. In current
4G networks, the UE regularly monitor the omnidirectional signal to
estimate the downlink channel. However, in 5G mmWave networks, this is difficult
to achieve due to the directionality and the rapid variations of the
channel \cite{giordani2018tutorial}. The directionality may significantly
delay the initial access procedure, especially for beams with narrow
beamwidth \cite{8407094}. To reduce the beam alignment overhead, efficient beam alignment algorithms
are therefore required. Motivated by this challenge, this paper proposes
a new beam alignment algorithm which exploits a noisy location information. 

Beam alignment has been previously studied
under the cell search phase \cite{6941329,7748573,8288065,7986409,7567506}
(i.e., the phase in which the UE search and connects to a BS
with a mutual agreement on transmission parameters). In \cite{7848981},
an exhaustive search algorithm to determine the optimal beam pair
was proposed. The exhaustive search is known for its high   complexity.
To reduce the delay in the exhaustive search algorithm, an efficient
hierarchical codebook adaptive algorithm is proposed in \cite{6600706}
to jointly search over the channel subspace. The proposed algorithm
allows the UE and BS to jointly align their beams within a constrained
time. In \cite{8376953}, a Bayesian tree search algorithm is proposed
to reduce the delay in the training of beamforming and combining vectors.

Beam alignment overhead can be reduced without compromising performance
if the location side information is available at both nodes. Indeed,
5G communication devices are expected to have access to location information
which can be obtained from GNSS satellites, sensors and 5G radio signal \cite{6924849,wymeersch20185g}.
In \cite{8313072}, exploiting location information for backhaul systems
was proposed. The authors showed that the time required for beam alignment
can be reduced if the position information is shared between
the nodes. For stationary backhaul systems, it is easy to assume that
perfect location information is available at both nodes since this
information can be obtained during installations. However, for a mobile
device, it is likely that the location information is noisy and designing
beam alignment algorithms might not be straightforward as in \cite{8313072}.
A noisy location information is considered in \cite{maschietti2017robust},
where the authors focused on an independent beam pre-selection at
the BS and the UE. While the pre-selection algorithm can improve the
beam selection speed, the performance of the selected beams cannot
be guaranteed since the beam selection decision is weighed on the
noisy location information. In addition, the decentralized beam selection
framework may degrade  system performance. 

This work focus on achieving fast and efficient transmit and receive beam alignment subject to a target rate constraint
by exploiting the noisy location information of the UE and potential reflecting points. The location information
of the static BS is assumed to be perfectly known, while the location
information of the reflecting points and UE are considered to be noisy.


The contributions of this paper are summarized as follows: 
Firstly, we propose a beam alignment algorithm iteratively executed at the BS and the UE
that exploits the location information of the UE and potential
reflecting points, this information is used to design a subset of beam codebook
from the BS and UE codebooks, thereby reducing the number of beam steering vectors
to be searched as compared to the exhaustive search algorithm. Secondly, we propose a search window in the  subset of beam codebooks to further reduce the angular space  to be searched. This is achieved by determining the direction of the actual beam after obtaining the local optimal beam within the search window. Furthermore, the local
optimal beam in each search is used to guide future beam search at the UE and BS thereby reducing the beam alignment overhead. Finally, we derive the Cram\'{e}r-Rao bound (CRB) of the channel parameter which an unbiased estimator should satisfy, the CRB is then used to model the channel estimation error.  We show by simulations that the proposed beam alignment scheme can  speed up the beam alignment process and reduce the beam alignment overhead when compared to existing schemes.

The rest of the paper is organized as follows. In  Section II, the signal model and the codebook used in this paper are introduced. Section III first describes the use of location information followed by a detail description of the proposed beam alignment algorithm. In Section IV, we present the  channel parameter estimation  and rate evaluation. Numerical results are shown in section V, and the paper is concluded in Section VI.

\subsection*{Notations}
Throughout this paper, matrices and vector symbols are represented
by uppercase and lowercase boldface respectively. $\mathbf{A}^{*}$,  $\mathbf{A}^{\mathsf{T}}$ and  $\mathbf{A}^{\mathsf{H}}$
represent the complex conjugate,  transpose and Hermitian transpose of the matrix $\mathbf{A}$ respectively. The mathematical expectation is denoted as $\mathbb{E}[.]$. tr$(\mathbf{A})$
represent the trace 
of matrix $\mathbf{A}$. The Kronecker product between two matrices  $\mathbf{A}$ and $\mathbf{B}$ is denoted as $\mathbf{A}\otimes\mathbf{B}$.

\begin{figure}
	\centering{}
	\psfrag{a}[bc][bc][1][0]{$[x_{\text{BS}},y_{\text{BS}}]$}
	\psfrag{b}[Bc][Bc][1][-40]{$\quad\quad[x_{1},y_{1}]$}
	\psfrag{c}[bc][bc][1][0]{$\quad[x_{2},y_{2}]$}
 	\psfrag{d}[bc][bc][1][-38]{$ [x_{\text{UE}},y_{\text{UE}}]$}
	\includegraphics[width=0.7\columnwidth]{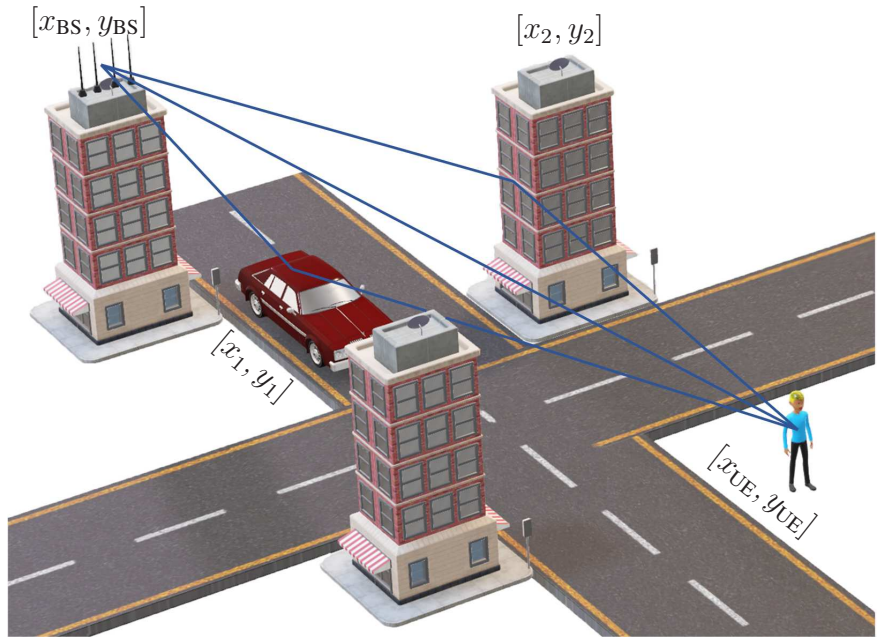}\caption{Example of network scenario with two reflectors \label{fig:Scenario}}
\end{figure}

\section{System Model}
 
Consider a wireless network scenario operating in the mmWave frequency
band and consisting of one UE, one BS, $M$ dominant paths with one line of sight (LOS) path and  $M-1$ reflected paths as shown in
Fig \ref{fig:Scenario}. The orientation of the UE is assumed to be fixed, however, the proposed scheme can also be extended to a scenario where the orientation of the UE is not known. In such scenario, the UE's orientation  can be estimated along with the location information \cite{8240645} after which the beam alignment algorithm proposed in this paper can be applied.  The Cartesian coordinate of each node defines its position while  the location of the BS is assumed to be known by the UE. The location information in this paper is discussed in  detail in the Section III. We assume that the UE and the BS are equipped with receive
and transmit uniform linear array (ULA) $N_{r}$ and $N_{t}$ antennas 
respectively. Furthermore, we assume the communication is made in
blocks (i.e., at discrete instance), where each block consists of $N$
slots. The channel is assumed to remain constant within each block
and change independently between blocks. From the $N$ slots, $N_a$
slots are used for the control phase within which beam alignment will be achieved, while $N-N_a$ slots are used for the data transmission phase. Note that the proposed beam alignment protocol requires both the uplink and downlink communication as the protocol is iteratively executed by both the BS and the UE. We assume the BS and UE are allowed to dwell in a slot with a fixed beamformer. More specifically, within each BS-slot,  the BS beamformer is kept fixed and used to transmit to the UE while  the UE  takes several measurements with different UE beam directions.  Similarly, in the UE-slot, the UE beamformer is fixed while the BS can take several measurements with  different BS beam directions.  We fix the number of beams that can be processed per slot as $N_b$. 
It can be observed that while data transmission can be improved by
selecting the best beam pair, the number of slots
$N_a$ taken to achieve beam alignment should be as low as possible
to reduce beam alignment overhead \cite{7880676,8407094}. Hence,
the proposed scheme aims to speed up the beam pair search subject to a rate constraint. 

\subsection{Signal and Channel  Model}
In this section, we present the signal model. 
We assume beamforming vector $\mathbf{v}_k $ where $k\in\{1,2,\ldots, N_t\}$ is employed at the BS while the UE employs  beamforming vector $\mathbf{u}_j$, where $ j\in\{1,2,\ldots, N_r\}$. Furthermore, we assume the beam vectors are normalized to unity: $\Vert\mathbf{v}_k\Vert  =\Vert\mathbf{u}_j\Vert=1$.  
The downlink received signal   can be expressed as   
\begin{equation}
y_{j,k}=\mathbf{u}_j^{\mathsf{H}}\mathbf{H}\mathbf{v}_ks+\mathbf{u}_j^{\mathsf{H}}\mathbf{n},
\end{equation}
where 
$\mathbf{u}_j\in\mathcal{C}^{N_{r}\times1}$  and $\mathbf{v}_k\in\mathcal{C}^{N_{t}\times1}$, $s$ is the transmitted symbol 
with unit energy $\left|s\right|=\sqrt{E_s}$, $E_s$ is the transmit energy,  $\mathbf{n}\in\mathcal{C}^{N_{r}\times1}$
is the complex Gaussian noise vector with  zero mean and covariance $\sigma^2 \mathbf{I}$.  
%
The downlink channel $\mathbf{H}$ is
expressed as \cite{6847111}
\begin{equation}
\mathbf{H}=\sqrt{N_tN_r}\sum_{m=0}^{M-1}\alpha_{m}\mathbf{a}_r(\theta_{m})\mathbf{a}_t^{\mathsf{H}}(\phi_{m}),\label{eq:uplinkchannel}
\end{equation}
where $M$ is the number of multi-paths, consisting of  one LOS path and $M-1$ reflected paths, specifically,  $m=0$ refers to the LOS path, and $m=1,\ldots,M-1$, refers to the $m$-th NLOS path passing through the $m$-th reflecting points. $\theta_{m}$ and $\phi_{m}$
are the angle of arrival (AOA) and angle of departure (AOD) of the $m$-th path at the receiver and transmitter in the downlink mode respectively, $\alpha_{m}\sim\mathcal{CN}(0,\sigma^{2}_{\alpha})$
denotes the instantaneous random complex gain for the $m$-th path.
The corresponding array response vectors at the UE and BS denoted as 
$\mathbf{a}_r(\theta_{m})$ and $\mathbf{a}_t(\phi_{m})$
are given by 
\setlength{\arraycolsep}{0.0em}
\begin{align}  
\mathbf{a}_r(\theta_{m})=\frac{1}{\sqrt{N_{r}}}\left[1,e^{-j\pi\cos\theta_{m}},\ldots, e^{-j\pi(N_{r}-1)\cos\theta_{m}}\right]^{\mathsf{T}},\label{eq:UE stering Vector}\\
\mathbf{a}_t(\phi_{m})=\frac{1}{\sqrt{N_{t}}}\left[1,e^{-j\pi\cos\phi_{m}},\ldots, e^{-j\pi(N_{t}-1)\cos\phi_{m}}\right]^{\mathsf{T}},\label{eq:BS stering Vector}
\end{align} respectively. 
We assume that the steering vectors are drawn from a codebook and the
design of the codebook is presented as follows.

\subsection{Codebook Structure}

The AOD and AOA are computed from the estimated location information
and are associated with  transmit and receive beamforming 
vectors respectively from a given codebook. The focus of this paper
is not the codebook design, therefore we refer our readers to \cite{7990158}
for details on codebook design. In this paper, the codebooks
are designed to  achieve approximately equal gain but with
narrow beams at the broadside and wide beams at the endfire \cite{7536855}.
The pointing angles at the UE and BS denoted as $\bar{\theta}_{j}$
and $\bar{\phi}_{k}$ respectively are separated into grids
as follows 
\begin{eqnarray}
\bar{\theta}_{j} =  \arccos\left(1-\frac{2\left(j-1\right)}{N_{r}-1}\right),\quad j=1, \ldots, N_{r},\label{eq:codebookrx}\\
\bar{\phi}_{k}  =  \arccos\left(1-\frac{2\left(k-1\right)}{N_{t}-1}\right),\quad k=1, \ldots, N_{t},\label{eq:codebooktx}
\end{eqnarray}
where $0\le\arccos\left(x\right)\le\pi$. The receive and transmit codebook is defined as 
\begin{eqnarray}
    \mathcal{U}&{}={}&\left[\mathbf{u}_1,\dots,\mathbf{u}_{N_r}\right],\\
    \mathcal{V}{}&{}={}&\left[\mathbf{v}_1,\dots,\mathbf{v}_{N_t}\right],
\end{eqnarray}
where $\mathbf{u}_j=\mathbf{a}_r(\bar{\theta}_j)$ and $\mathbf{v}_k=\mathbf{a}_t(\bar{\phi}_k)$  are the beam steering vectors over the discrete grid angles.  It can be observed from
(\ref{eq:codebookrx}) and (\ref{eq:codebooktx}), that the codebook
covers a large angular space. 
 The exhaustive search requires that the BS and UE
search through the entire codebook $\mathcal{U}$ and $\mathcal{V}$
respectively. The beam alignment is achieved by selecting the beam pair that maximizes
the downlink rate, which can be mathematically expressed as 
\begin{equation}
\underset{\left\{{\mathbf{u},\mathbf{v}}\right\}\in\mathcal{U}\times\mathcal{V}}{\max}R_d(\mathbf{u},\mathbf{v}),\label{eq:optimal beams}
\end{equation} where 
 \begin{equation}
 R_d\left(\mathbf{u},\mathbf{v}\right)=\log_{2}\left(1+\frac{E_s \left|\mathbf{u}^{\mathsf{H}}\mathbf{H}\mathbf{v}\right|^{2}}{\sigma^2}\right).\label{eq:rate}
 \end{equation}
 Note that in (\ref{eq:optimal beams}), we assume  that the channel state information is perfectly known, which enables us the know the SNR at the receiver and evaluate average rate  under any choice of $\mathbf{v}$  and $\mathbf{u}$.  However, performing beam alignment by exhaustive search method may
incur high system overhead. In addition, due to the dynamic nature
of the channel, especially in a mobile scenario this method may not be suitable for 5G communication.   

Hence, we focus on achieving fast transmit and receive beam alignment subject to a target rate $R_0$, where $R_0\le R_d\left(\mathbf{u},\mathbf{v}\right)$. The proposed scheme is discussed in details in the following section.

\section{Beam Alignment with Location Information}

In this section, we focus on the proposed beam alignment  algorithm. We assume that
the location information of possible reflecting points can be independently
estimated by the BS and the UE \cite{8240645,3gpp}.  However, we note
that the location information could be erroneous, and the uncertainty
of the location information at the nodes are included in the design
of the proposed algorithm.

\subsection{Exploiting Location Information} \label{sec:location information with error}


Location information  can be obtained with the use of available positioning technologies. 
The location information obtained at the BS and UE may be noisy due to latency in the position information exchange or due to the use of different positioning technologies at the UE and BS. For instance, the BS may be able to estimate potential reflecting points more
accurately than the UE due to interactions with multiple UEs. 

We define the location matrix $\mathbf{L}_i\in \mathbb{R}^{2\times M}$ containing the actual location coordinate of the nodes, where the nodes refer to the potential reflecting points and either the BS or UE. Specifically, when the location information is estimated from the BS, the location information matrix contains the coordinate information of the UE and the reflecting points. Similarly, when the location information is measured from the  UE, the location information matrix contains the coordinate information of the BS and the reflecting points. Hence,  we express $\mathbf{L}_i$ as follows
\begin{equation}
    \mathbf{L}_{i}=\left[\mathbf{l}_{0},\mathbf{l}_{1},\ldots,\mathbf{l}_{{M-1}}\right], \quad \label{eq:location information} 
    \end{equation}
    where 
    \begin{equation}
        \mathbf{l}_{0}=\begin{cases}
    \mathbf{l}_{\text{BS}}\quad \text{for } i=\text{UE},\\
    \mathbf{l}_{\text{UE}}\quad \text{for } i=\text{BS},
    \end{cases}
    \end{equation}
and $\mathbf{l}_{m}=[x_m,y_m]^{\mathsf{T}}$ is the location information of the node along path $m$.   The location information of the observer is denoted as $\mathbf{l}_{\text{BS}}$ for the BS and $\mathbf{l}_{\text{UE}}$ for the UE.
Note from (\ref{eq:location information}) that when the UE is the observer, $\mathbf{l}_{0}=\mathbf{l}_{\text{BS}}$, and when the BS is the observer,  $\mathbf{l}_{0}=\mathbf{l}_{\text{UE}}$.
Furthermore,  we model the independent location information of the UE and  $M-1$ reflecting points available at the BS  as 
\begin{eqnarray}
\hat{\mathbf{L}}_{\text{BS}}=\mathbf{L}_{\text{BS}}+\mathbf{E}_{\text{BS}},
\label{eq:BS estimate}
\end{eqnarray}
 where $\mathbf{E}_{\text{BS}}$ is 
 the matrices containing the random location estimation errors of the $x$ and $y$ coordinates made by the BS 
 given as 
 \begin{eqnarray}
 \mathbf{E}_{\text{BS}}=[\mathbf{e}_{0}^{\text{BS}},\mathbf{e}_{1}^{\text{BS}},\ldots,\mathbf{e}_{{M-1}}^{\text{BS}}],\label{BS location erro}
 \end{eqnarray}
 where the superscript BS and UE are used to indicate the observation at the BS and UE respectively.
In this paper, we adopt a uniform bounded error model for the location estimation error \cite{maschietti2017robust,7536855}. We assume that all the estimates lie within a disk centered on the estimated location. Let $S(r_m)$ 
be the two-dimensional disk centered at the estimated  location of the node in path $m$ with radius $r_m$. Here, we refer to the disk with radius $r_m$ as the uncertainty region in path $m$. The random estimation  error $\mathbf{e}_{m}^{\text{BS}}$ is uniformly distributed in $S\left(r_m^{\text{BS}}\right)$ in (\ref{BS location erro}), 
 such that $r_m^{\text{BS}}$ is 
 the maximum position error of the node in path $m$ as seen from the BS. In this paper, we assume that the  location information of the BS is perfectly known. In addition,  when the UE is the observer, $r_{0}^{\text{UE}}=r_{\text{BS}}^{\text{UE}}$, and when the BS is the observer,  $r_{0}^{\text{BS}}=r_{\text{UE}}^{\text{BS}}$, where $r_{\text{BS}}^{\text{UE}}$ is the maximum position error of the BS observed at the UE  and and $r_{\text{UE}}^{\text{BS}}$ is the maximum position error of the UE observed at the BS. On the other hand, we assume that the location of the UE could contain some uncertainty when observed by the UE itself, hence, we denote the maximum position error of the UE when observed by the UE itself as $r_{\text{UE}}^{\text{UE}}$. For ease of notation, we denote the vector $\mathbf{r}_{i}=\left[r_0^i,r_1^i,\ldots,r_{M-1}^i\right]$ containing the maximum location errors of the nodes observed at the UE and BS for $i\in\{\text{UE},\text{BS}\}$.  
From the   estimated location information at the BS (i.e., $\hat{\mathbf{L}}_{\text{BS}}$), the AOD of the $m$-th path can be computed as 
\begin{equation}
\hat\phi_{m}=\frac{\pi}{2}-\arctan\left(\frac{\hat{x}_{m}^{\text{BS}}-x_{\text{BS}}}{\hat{y}_{m}^{\text{BS}}-y_{\text{BS}}}\right), \quad m=0,\ldots,M-1, \label{eq:AOD}
\end{equation} 
  
Similarly, the independent location information of the BS  and $M-1$ reflecting points available at the UE can be modelled as (\ref{eq:BS estimate}) and (\ref{BS location erro}), where the random estimation  error $\mathbf{e}_{m}^{\text{UE}}$ from the UE is uniformly distributed in $S\left(r_m^{\text{UE}}\right)$.
Following a similar procedure in (\ref{eq:BS estimate}) to (\ref{eq:AOD}), the AOA  of the $m$-th path to the UE can be evaluated from the location information available  at the UE as
\begin{equation}
\hat\theta_{m}=\frac{\pi}{2}-\arctan\left(\frac{\hat{x}_{m}^{\text{UE}}-\hat{x}_{\text{UE}}}{\hat{y}_{m}^{\text{UE}}-\hat{y}_{\text{UE}}}\right), 
\quad  m=0,\ldots,M-1,
\label{eq:AOA}
\end{equation} 
 where $\hat{x}_{\text{UE}}$ and  $\hat{y}_{\text{UE}}$  are the estimated $x$ and $y$ coordinate of the UE since its position is uncertain. 

The estimated distance information at the BS and UE can be obtained from $\hat{\mathbf{L}}_{\text{BS}}$ and $\hat{\mathbf{L}}_{\text{UE}}$ respectively as
\begin{eqnarray}
\hat{d}_{m}^{\text{BS}}  =  \sqrt{\left(x_{\text{BS}}-\hat{x}_{m}^{\text{BS}}\right)^{2}+\left(y_{\text{BS}}-\hat{y}_{m}^{\text{BS}}\right)^{2}}\label{eq:BSdistance},\\ 
\hat{d}_{m}^{\text{UE}}  =  \sqrt{\left(\hat{x}_{\text{UE}}-\hat{x}_{m}^{\text{UE}}\right)^{2}+\left(\hat{y}_{\text{UE}}-\hat{y}_{m}^{\text{UE}}\right)^{2}}\label{eq:UEdistance}. 
\end{eqnarray}

\subsection{Proposed Coordinated Beam Alignment }

We aim to reduce the search overhead by taking advantage of the
location information while accounting for the location estimation
error. We assume that the BS is located at the origin and its location
is perfectly known by the UE.  Based on the estimated location information and  maximum location information error, we propose a coordinated beam alignment algorithm with
beams subsets at  the BS and UE denoted as $\mathcal{B}_{\text{BS}}^m\subset \mathcal{V}$
and $\mathcal{B}_{\text{UE}}^m\subset \mathcal{U}$ respectively.
\begin{figure}
	
	\centering{}
	\psfrag{a}[bc][bl][1][0]{$\quad\quad\quad\quad[x_{\text{BS}},y_{\text{BS}}]$}
	\psfrag{b}[bl][bl][1][0]{$r_1^{\text{BS}}$}
	\psfrag{c}[bc][bc][1][0]{$\quad\quad\quad\left[\hat{x}_1^{\text{BS}},\hat{y}_1^{\text{BS}}\right]$}
	\psfrag{d}[bl][bl][1][0]{$r_{0}^{\text{BS}}$}
	\psfrag{e}[bc][bl][1][0]{$\quad\quad\:\:\left[\hat{x}_{0}^{\text{BS}},\hat{y}_{0}^{\text{BS}}\right]$}
	\psfrag{eu}[bl][bl][1][0]{$\hat{\phi}_{m}-\epsilon_m$}
	\psfrag{el}[bl][bl][1][0]{$\hat{\phi}_{m}+\epsilon_m$}
	\psfrag{f}[bl][bl][1][0]{Uncertainty region}
	\psfrag{g}[bl][bl][1][0]{$\hat{\phi}_m$}
	\psfrag{h}[bl][bl][1][0]{${\gamma}_m$}
	\psfrag{x}[bl][bl][1][0]{$x$}
	\psfrag{y}[bc][bl][1][0]{$\quad y$}
	\includegraphics[width=0.7\columnwidth]{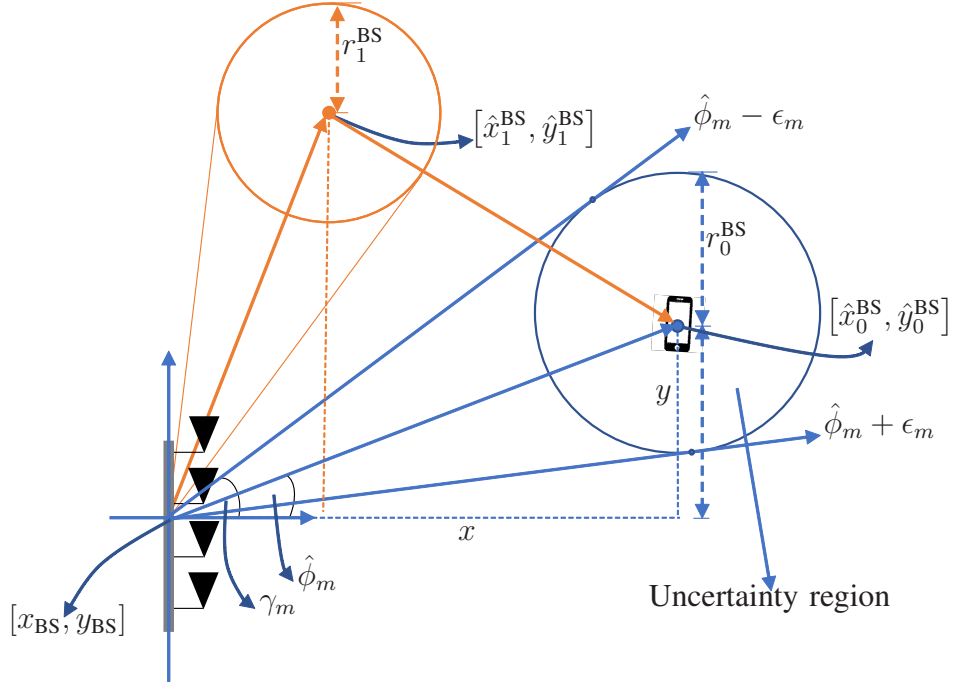}\caption{Example  of two-dimensional system model showing the estimated location of the UE, a reflecting point and the uncertainty region measured by the BS. \label{fig:Error-boundary} }

\end{figure}

\subsubsection{Construction of the Codebook Subset $\mathcal{B}^m_{\text{BS}}$ and $\mathcal{B}^m_{\text{UE}}$}

%


Define \begin{equation}
f_{\text{BS}}(\phi)= \underset{\mathbf{v}\in\mathcal{V}} \argmin\left\Vert \mathbf{v}-\mathbf{a}_t({{\phi}})\right\Vert _{F}^{2},\label{eq:Fv_j}
\end{equation}
\begin{equation}
f_{\text{UE}}(\theta)=  \underset{\mathbf{u}\in\mathcal{U}}\argmin\left\Vert \mathbf{u}-\mathbf{a}_r({{\theta}})\right\Vert _{F}^{2},\label{eq:Fu_j}
\end{equation}
as the functions that return the closest beam vectors to $\phi$  and $\theta$ respectively. Then the optimal rate can be obtained by searching through  a subset of beam vectors in the uncertainty regions  which is  summarized in the following proposition.  
\begin{prop}
The optimal rate can be recovered from solving the following reduced search problem within  the uncertainty region as
\begin{equation}
	\underset{m}{\max}\underset{\left(\mathbf{u},\mathbf{v}\right)\in\mathcal{B}_{\text{BS}}^{m}\times\mathcal{B}_{\text{UE}}^{m}}{\max}R_d\left(\mathbf{u},\mathbf{v}\right),
	\label{optimalmaxrate}
\end{equation} 
where 
\setlength{\arraycolsep}{0.0em}
\begin{eqnarray}
\mathcal{B}^{m}_{\text{BS}} &{} \triangleq{}& \left[ f_{\text{BS}}(\bar{{\phi}}_k): \hat{\phi}_m -  \epsilon_m \leq \bar{{\phi}}_k \leq \hat{\phi}_m +  \epsilon_m\right],\label{eq:BBS} \forall{k},\\  \mathcal{B}^{m}_{\text{UE}}  &{}\triangleq{}& \left[ f_{\text{UE}}(\bar{\theta}_j): \hat{\theta}_m -  \delta_m \leq \bar{\theta}_j \leq \hat{\theta}_m +  \delta_m\right], \forall{j},\label{eq:BUE}
\end{eqnarray} 
are the subset of beam vectors that lies within the error boundary of the $m$-th path, $\bar{\theta}_j$ and $\bar{\phi}_k$ are given by (\ref{eq:codebookrx}) and (\ref{eq:codebooktx}) respectively, the parameters $\epsilon_m$ and $\delta_m$ are given by 
\begin{eqnarray}
\epsilon_m=  \arctan\left(\frac{\hat{d}_{m}^{\text{BS}}\sin \hat{\phi}_{m} +r_{m}^{\text{BS}}}{\hat{d}_{m}^{\text{BS}}\cos\hat{\phi}_{m}}\right)-\hat{\phi}_m,\label{transmitter subset} 
\\
\delta_m  = \arctan\left(\frac{\hat{d}_{m}^{\text{UE}}\sin\hat{\theta}_{m}+r_{m}^{\text{UE}}}{\hat{d}_{m}^{\text{UE}}\cos\hat{\theta}_{m}}\right)-\hat{\theta}_m, 
\end{eqnarray}
respectively, the derivations of $\epsilon_m$ and $\delta_m$ are presented in Appendix \ref{proof proposition 1}.

\end{prop}

\begin{rem}

	From Proposition 1, it can be observed that the number of beams in the set $\mathcal{B}^m_{\text{UE}}$ and $\mathcal{B}^m_{\text{BS}}$ decreases as the maximum location error tends to zero (i.e., with high precision in the location information). Furthermore, the fewer beam steering vectors in  $\mathcal{B}^m_{\text{UE}}$ and $\mathcal{B}^m_{\text{BS}}$  enable
	fast beam alignment as compared to exhaustively searching over the
	entire angular space. 
\end{rem}

While the optimal rate can be obtained from (\ref{optimalmaxrate}), in what follows, we present a low complexity beam alignment procedure to speed up the beam search subject to a target rate constraint $R_0$.

\subsubsection{Design of the Search Window}

\begin{figure}
	\centering{}
	\psfrag{A}[Bl][bl][1][0]{Optimal beam direction}
	\psfrag{B}[bt][bl][1][0]{$\mathcal{B}_{\text{BS}}^m$}
	\psfrag{D}[bl][bl][1][0]{Direction of search}
	\psfrag{E}[bc][bl][1][0]{Estimated beam direction}
	\psfrag{F}[bc][bl][1][0]{First search}
	\psfrag{N}[bc][bl][1][0]{Next search}
	\psfrag{S}[bc][br][1][0]{Search window}
	\psfrag{W}[bc][bl][1][0]{$\mathcal{W}_{\text{BS}}^m$}
	\psfrag{L}[Bl][bl][1][0]{Local optimal beam in search window}
	
	\includegraphics[width=0.5\columnwidth]{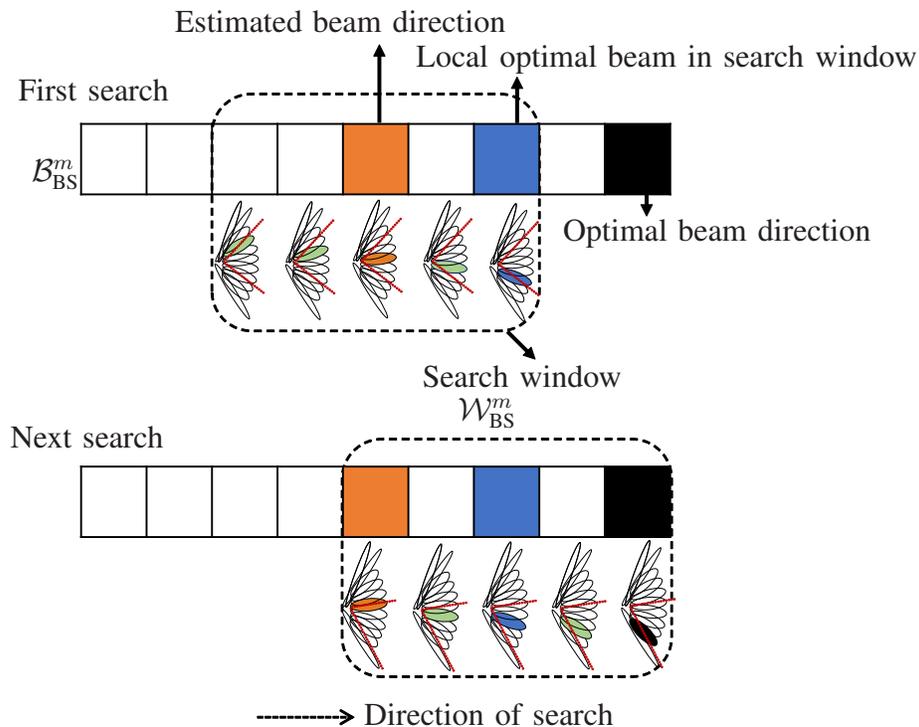}\caption{Example of proposed scenario showing  two successive beam pointing vector search in $\mathcal{B}_{\text{BS}}^m$ with a search window of $W_{\text{BS}}^m=5$ beam vectors, where the local optimal beam determines the direction of search, hence the two pointing vectors on the left are not searched. \label{fig:Example-of-beam}}
\end{figure}

As discussed in the previous section, by exploiting the location information, we can limit the number
of beam vectors in the codebook $\mathcal{V}$ and $\mathcal{U}$, hence, we obtained the new  beam subsets $\mathcal{B}^m_{\text{BS}}$ and $\mathcal{B}^m_{\text{UE}}$ for each of the $m$-th path. Instead of sweeping through the entire beam vectors in $\mathcal{B}^m_{\text{BS}}$ and $\mathcal{B}^m_{\text{UE}}$, we can further reduce the number of beam vectors to be searched by measuring across a small window of beams $\mathcal{W}_{\text{BS}}$ and $\mathcal{W}_{\text{UE}}$ at the
BS and UE respectively to determine the direction of beam search as
shown in Fig \ref{fig:Example-of-beam}.
Let the size of the window at the BS (resp. UE) be denoted as $W_{\text{BS}}=\left|\mathcal{W}_{\text{BS}}\right|$
(resp. $W_{\text{UE}}=\left|\mathcal{W}_{\text{UE}}\right|$), where $W_{\text{BS}}$, $W_{\text{UE}}$ are jointly determined at the BS and UE and correspond to the number of beams that can be processed in a slot $N_b$.  
Furthermore, let the index of the estimated beam in $\mathcal{B}_{\text{BS}}^m$ (resp. $\mathcal{B}_{\text{UE}}^m$) be denoted as $I_{\text{BS}}^m$ (resp. $I_{\text{UE}}^m$), then the set of beam vectors in $\mathcal{W}_{\text{BS}}$  and $\mathcal{W}_{\text{UE}}$ are defined as follows  
\begin{eqnarray}
	\mathcal{W}_{\text{BS}}&{}={}&\begin{cases}
	\mathcal{B}_{\text{BS}} & \text{for }  W_{\text{BS}}\ge B_{\text{BS}},\\
	\left[
	\mathbf{v}_{B_{\text{BS}}-W_{\text{BS}}+1},\ldots, \mathbf{v}_{B_{\text{BS}}}\right] & \text{for }W_{\text{BS}}<B_{\text{BS}}\text{ and }B_{\text{BS}}-I_{\text{BS}}<W_{\text{BS}}/2,\\
	\left[
	\mathbf{v}_1,\ldots, \mathbf{v}_{W_{\text{BS}}}\right] & \text{for }W_{\text{BS}}<B_{\text{BS}}\text{ and }I_{\text{BS}}<W_{\text{BS}}/2,\\
	\left[
	\mathbf{v}_{I_{\text{BS}}-\lfloor W_{\text{BS}}/2 \rfloor},\ldots,\mathbf{v}_{I_{\text{BS}}+\lfloor W_{\text{BS}}/2\rfloor}\right] & \text{for }W_{\text{BS}}<B_{\text{BS}}\text{ and }I_{\text{BS}}>W_{\text{BS}}/2,
	\end{cases}\label{eq:window}\\
	\mathcal{W}_{\text{UE}}&{}={}&\begin{cases}
	\mathcal{B}_{\text{UE}} & \text{for }  W_{\text{UE}}\ge B_{\text{UE}},\\
	\left[
	\mathbf{v}_{B_{\text{UE}}-W_{\text{UE}}+1},\ldots, \mathbf{v}_{B_{\text{UE}}}\right] & \text{for }W_{\text{UE}}<B_{\text{UE}}\text{ and }B_{\text{UE}}-I_{\text{UE}}<W_{\text{UE}}/2,\\
	\left[
	\mathbf{v}_1,\ldots, \mathbf{v}_{W_{\text{UE}}}\right] & \text{for }W_{\text{UE}}<B_{\text{UE}}\text{ and }I_{\text{UE}}<W_{\text{UE}}/2,\\
	\left[
	\mathbf{v}_{I_{\text{UE}}-\lfloor W_{\text{UE}}/2 \rfloor},\ldots,\mathbf{v}_{I_{\text{UE}}+\lfloor W_{\text{UE}}/2\rfloor}\right] & \text{for }W_{\text{UE}}<B_{\text{UE}}\text{ and }I_{\text{UE}}>W_{\text{UE}}/2,\\
	\end{cases}\label{eq:UEwindow}
	\end{eqnarray}
where $B_{\text{BS}}=\left|\mathcal{B}_{\text{BS}}\right|$, $B_{\text{UE}}=\left|\mathcal{B}_{\text{UE}}\right|$ and
$\lfloor(a/b)\rfloor$ denotes the floor of the operation. Note that since the design of the window is similar for each path, the index $m$ is dropped for ease of notation in (\ref{eq:window}) and (\ref{eq:UEwindow}).  

When the  size of the window is larger or equal to the size of the beam set, the beam vectors in the window are given by the the beam vectors in the  beam set.
If the size of the beam set  is larger than the size of the window (i.e., $B_{\text{BS}}> W_{\text{BS}}$),  the range of beam vectors in the search window is 
defined by   (\ref{eq:window}), where the center of the window is determined by the position of the estimated beam vector. The implementation of the window is discussed in  the proposed beam alignment procedure  in the following section.
\subsubsection{Low Complexity Beam Alignment\label{beam alignement}}

The process begins with the acquisition of the location information. The BS estimates the noisy location information of the UE and the possible reflecting points at $n=1$ as discussed in Section \ref{sec:location information with error}, while the UE estimates its own location information and the location information of  the reflecting points. Thereafter,  the BS and UE evaluate the AODs and AOAs   for each  path respectively. We assume that the BS and UE agree on the ordering of the paths based on the angle information (i.e., based on the computed AOA and AOD), where $m=0$ denotes the LOS path and $m=1,\ldots,M-1$ denote the path through the first reflecting point to the $M-1$ reflecting points respectively.

In the beam alignment phase, (i.e., $2\le n\le N_a$), the BS and UE jointly search the beam vectors in the uncertainty region corresponding to each of the $m$-th path to determine the pair of beam vectors that satisfies the target rate $R_0$. Specifically, in the downlink mode,  the BS selects the beam steering vector $\mathbf{v}_{m}$ from $\mathcal{B}^m_{\text{BS}}$ which is closest to the computed AOD $\hat{\phi}_m$ and transmit to the UE, such that 
\begin{equation}
    \mathbf{v}_{m} = f_{\text{BS}}(\hat{\phi}_m),\label{eq:v_j}
\end{equation} 
while the UE selects a combining vector $\mathbf{u}_{m}$ 
which correspond to the  computed AOA from $\mathcal{B}^m_{\text{UE}}$ such that
\begin{equation}
\mathbf{u}_{m} = f_{\text{UE}}(\hat{\theta}_m).\label{eq:u_m}
\end{equation}
The observed rate from path $m$ i.e., $R_{d,m}(\mathbf{u}_m,\mathbf{v}_m)$ is compared with a threshold rate $R_{0}$, if the target $R_{0}$ is met, the UE sends a message to the BS to move to data transmission phase. Note that $R_{d,m}\left(\mathbf{u},\mathbf{v}\right)$ is the rate  obtained in path $m$ which can be evaluated as (\ref{eq:rate}). On the hand, if the target rate is not satisfied, the transmit beamforming vector at the BS is kept fixed while the UE takes several measurements within a small window of beam set $\mathcal{W}^m_{\text{UE}}\subset\mathcal{B}^m_{\text{UE}}$ to determine the direction of beam search in $\mathcal{B}^m_{\text{UE}}$. 
At the end of the UE beam measurement, a local optimal beam
which maximizes the rate in the window is selected at the UE. If the local optimal beam satisfies the target rate requirement, the beamforming vector from the BS and the local optimal beam are said to be aligned and the UE sends a message to the BS to begin data transmission,  we refer to this phase as the UE beam alignment phase. If the target rate is not met at the end of the UE beam alignment phase, the UE sends a message to the BS to continue with beam alignment.  The objective at the UE in each UE beam alignment phase can be mathematically expressed as
\begin{equation}
\mathbf{u}_{\text{op}} =\underset{\mathbf{u}\in\mathcal{W}^m_{\text{UE}}}\argmax \:R_{d,m}\left(\mathbf{u},\mathbf{v}_{m}\right).\label{eq:Rate at MS}
\end{equation}

In the uplink mode, the  UE transmits  to the BS with the local optimal beam $\mathbf{u}_{\text{op}}$ obtained from the UE beam alignment phase where we refer to this phase as the BS beam alignment phase. In this phase, the local optimal beam at the UE is kept fixed, while the BS
takes several measurements from a small window of beam vectors $\mathcal{W}^m_{\text{BS}}\subset\mathcal{B}^m_{\text{BS}}$
to determine the direction of search in $\mathcal{B}^m_{\text{BS}}$  as shown in Fig. \ref{fig:Example-of-beam}. At the end of the measurements, a local optimal beam is selected at the BS, this local optimal  beam is used   to transmit to the UE for the next UE beam alignment phase. The objective in each BS beam alignment phase can be expressed as 
\begin{equation}
\mathbf{v}_{\text{op}}=\underset{\mathbf{v}\in\mathcal{W}^m_{\text{BS}}}\argmax \:R_{u,m}\left(\mathbf{v},\mathbf{u}_{\text{op}}\right),\label{eq:Rate at BS}
\end{equation}
where the uplink rate can be evaluated as  
\begin{equation}
 R_{u,m}\left(\mathbf{v},\mathbf{u}_{\text{op}}\right)=\log_{2}\left(1+\frac{E_s\left|\mathbf{v}^{\mathsf{T}}\mathbf{H}^{\mathsf{T}}\mathbf{u}^*_{\text{op}}\right|^{2}}{\sigma^2}\right),\label{eq:uplinkrate}
 \end{equation}
 where $\mathbf{H}^{\mathsf{T}}$ is the uplink channel. 
In the subsequent UE and BS alignment phase, the search is performed over the local optimal beam obtained from the previous beam search as shown in Fig \ref{fig:Example-of-beam}, (i.e., the window is centered on the local optimal beam direction obtained from the previous search).  The BS and UE alternately perform beam alignment based on the procedure discussed above for path $m$. 

The beams are said to be aligned if 
\begin{equation}
 R_{d,m}\left(\mathbf{u}_{\text{op}},\mathbf{v}_{\text{op}}\right)\ge R_0,\label{eq: optimal rate}
 \end{equation} 
 is satisfied. If the   objective in (\ref{eq: optimal rate}) is not satisfied for path $m$, the search is performed for path $m+1$. The process is repeated for each path until   (\ref{eq: optimal rate}) is satisfied or the scan over all the paths are carried out.
 A summary of the proposed coordinated beam alignment is presented in Algorithm  \ref{Algorithm-I-Coordinated}.

\begin{rem}

From the proposed algorithm, it can be observed that for
a given BS beamforming vector in each downlink mode, a local optimal receive beam vector
can be obtained at the UE. The local optimal beam vector can be used
to transmit  to the BS in the uplink mode to guide the selection of beamforming
vector during the BS beam alignment phase. In addition, determining the search direction from the search window can further reduce the
number of beam vectors to be searched and thereby speed up the beam alignment process.

\end{rem}

\begin{algorithm}[hpbt!]
	\caption{Coordinated Beam Alignment in one block}
 \label{Algorithm-I-Coordinated}
	\begin{algorithmic}[1]
		\Statex \textbf{Input:}  $\mathbf{L}_i$, 
		$N$, $W_{\text{BS}}$, $W_{\text{UE}}$,
		$\mathbf{r}_{\text{UE}}$, $\mathbf{r}_{\text{BS}}$
		\Statex \textbf{Output:}  $\mathbf{u}_{\text{op}},\:\mathbf{v}_{\text{op}},\:N_a$
		\Statex 
		\For {$n=1,2,\ldots, N$}
		\If {$n=1$}
		\State {{Estimate $\hat{\mathbf{L}}_{\text{BS}}$ and $\hat{\mathbf{L}}_{\text{UE}}$}}
		\EndIf
		
		\State Evaluate ${\hat{\phi}}_{m}$ and ${\hat{\theta}}_{m} \quad \forall{m}$ from
		(\ref{eq:AOD}) and (\ref{eq:AOA})
		\State $n\leftarrow n+1$
	
		\For {$m=0,1,\dots,M-1$}
		\State {Select $\mathbf{v}_m$  and $\mathbf{u}_m$} from (\ref{eq:v_j})  and (\ref{eq:u_m}) respectively.
		\If {$R_{d,m}\left(\mathbf{u}_{m},\:\mathbf{v}_{m}\right)\ge R_{0}$\label{BS alignment}
		}
		\State Begin data transmission phase (line \ref{data})
		\Else
		\parState{%
		Construct a finite beam vectors set 
		$\mathcal{B}^m_{\text{UE}}$ according to (\ref{eq:BUE})}
		\State Set the window according to (\ref{eq:window})
		\parState{Determine $\mathbf{u}_{\text{op}}$ from (\ref{eq:Rate at MS}) and transmit to BS}
		\State $n\leftarrow n+1$ 
		\EndIf
		
		\If {$R_{u,m}\left(\mathbf{v},\mathbf{u}_{\text{op}}\right)\ge R_{0}$}
		\State Go to line \ref{data} 
		\Else
		\parState{Construct a finite beam vectors set $ \mathcal{B}^m_{\text{BS}}$ according to (\ref{eq:BBS})}
		\State Set the window according to (\ref{eq:window})
		\parState{Determine $\mathbf{v}_{\text{op}}$ according to (\ref{eq:Rate at BS}) and transmit to UE}
		\State $n\leftarrow n+1$ 
		\State Return to line \ref{BS alignment}
		\EndIf
		\EndFor
		
		\State Output  $\mathbf{v}_{\text{op}}$,  $\mathbf{u}_{\text{op}}$ and $N_a=n$ \label{data}
		\State Evaluate performance using (\ref{eq:penalty})	
		
		\EndFor	
		
	\end{algorithmic}
\end{algorithm}
 At the end of the beam alignment process, the UE and BS move to data transmission phase where we evaluate the effective rate as follows
\begin{equation}
R_{\text{eff}}=\left[\left(1-\frac{N_a}{N}\right)R_{d,m}\left(\mathbf{u}_{\text{op}},\:\mathbf{v}_{\text{op}}\right)\right]^+,\label{eq:penalty}
\end{equation}
where $(N-N_a)/N$ is the fraction of the slots allocated to
data transmission and $\left[a\right]^{+}$ denotes $\max\:(a,0)$. For the exhaustive search algorithm, the parameter $N_a$ can be evaluated as $N_a=\lceil N_tN_r/N_b\rceil$, where $N_b$ is the number of beams processed in each  slot, $N_r$ and $N_t$ correspond to the number of beams at the UE and BS respectively as given by (\ref{eq:codebookrx}) and (\ref{eq:codebooktx}).


\section{Cram\'er-Rao  Bounds for Channel Parameters Estimation} 

In the previous sections, it is assumed that the  channel matrices and the channel state information are perfectly known. Unfortunately, this assumption may not be true in real-world transmission but have to be obtained by channel estimation algorithms \cite{6847111}. As a consequence, a more or less accurate estimate of the channel is available for the rate computations.  Hence, we estimate 
the AOA, AOD and $\alpha_m$  using the Cram\'er-Rao Bound \cite{stoica1989music} and the estimates are used to reconstruct the channel.  We focus on the downlink channel estimation where the UE takes  measurements in different spatial direction and sends feedback messages to the BS using the local optimal beam $\mathbf{u}_{\text{op}}$ as discussed in the previous section.

In general, the channel matrix estimate  $\tilde{\mathbf{H}}$  can be considered as the sum of the true channel matrix $\mathbf{H}$ and the channel estimation error matrix $\mathbf{E}$ which can be expressed as 
\setcounter{equation}{34}
\begin{equation}
    \tilde{\mathbf{H}}=\mathbf{H}+\mathbf{E}.
\end{equation}
 Since the channel estimation error matrix $\mathbf{E}$ is not known, we treat the elements of  $\mathbf{E}$ as random variables such that
$\mathbf{E}\sim\mathcal{CN} (\mathbf{0}, \boldsymbol{\Sigma}_{\mathbf{H}})$, where  $\boldsymbol{\Sigma}_{\mathbf{H}}$ is the covariance matrix obtained from the Fisher information matrix which will be discussed subsequently.
In what follows, we derive the Fisher information matrix.    
\subsection{Likelihood Function}

Let $P$ and $Q$ denote the sum of the number of beam  steering vectors in the uncertainty regions  at the BS and UE respectively, where $P=\sum_mB_{\text{BS}}^m$ and $Q=\sum_mB_{\text{UE}}^m$. 
If each of the $P$ transmit beamforming vectors 
are measured against each of the $Q$ receive beamforming vectors, the observation $\mathbf{Y} \in \mathbb{C}^{P \times Q}$ can be expressed in the form
of a matrix given by  
\begin{equation}
   \mathbf{Y}=\left[ 
      \mathbf{y}_{1},  \mathbf{y}_{2},  \ldots, \mathbf{y}_{Q}\right], 
\end{equation}
 where 
  $\mathbf{y}_{q}=\left[ 
      y_{1,q}, y_{2,q}, \ldots,  y_{P,q}\right]^{\mathsf{T}}$.

Let 
\begin{align}
\mathbf{y}=\operatorname{vec}(\mathbf{Y})
=\left[
      \mathbf{y}_{1}^{\mathsf{T}}, \mathbf{y}_{2}^{\mathsf{T}}, \ldots,  \mathbf{y}_{Q}^{\mathsf{T}} \right]^{\mathsf{T}}, 
      \end{align}
then 
\setlength{\arraycolsep}{0.0em}
\begin{eqnarray}
\mathbf{y} &{}={}& \operatorname{vec}\left(\mathbf{U}^{\mathsf{H}}\mathbf{H}\mathbf{V}\right)+\operatorname{vec}\left(\mathbf{N}\right)\nonumber \\
&{}={}&  \sum_{m=0}^{M-1}\left(\underset{\mathbf{A}}{\underbrace{\mathbf{V}^{\mathsf{T}}\otimes\mathbf{U}^{\mathsf{H}}}}\right)\operatorname{vec}\left(\mathbf{H}\right)+\mathbf{n} \nonumber \\
&{}={}& \sum_{m=0}^{M-1}\mathbf{A}\left(\mathbf{a}_t^{*}(\phi_{m})\otimes\mathbf{a}_r^{\mathsf{H}}(\theta_{m})\right)+\mathbf{n},\label{eq:signal vector}
\end{eqnarray} 
where $\mathbf{U}\triangleq\left[\mathbf{u}_{1},\mathbf{u}_{2},\ldots,\mathbf{u}_{P}\right]$,
$\mathbf{V}\triangleq\left[\mathbf{v}_{1},\mathbf{v}_{2},\ldots,\mathbf{v}_{Q}\right]$,
 $\mathbf{N}$ denotes a complex matrix whose entries are assumed to be uncorrelated, each with zero mean and variance $\sigma^2$ and  $\mathbf{n}\sim \mathcal{CN}\left(\mathbf{0},\sigma^{2}\mathbf{I}_{PQ}\right).$
The mean vector can be expressed as 
\begin{eqnarray}
\mathbf{m}  =  \sum_{m=0}^{M-1}\mathbf{A}\left(\mathbf{a}_t^{*}(\phi_{m})\otimes\mathbf{a}_r^{\mathsf{H}}(\theta_{m})\right).\label{eq:mean}
\end{eqnarray}
Let $\boldsymbol{\theta}=\left[\phi_{0},\ldots,\phi_{M-1}, \theta_{0},\ldots, \theta_{M-1}, \alpha_{0},  \ldots,\alpha_{M-1}\right]^{\mathsf{T}}$,
then the likelihood function of the random
vector in  (\ref{eq:signal vector}) conditioned on $\boldsymbol{\theta}$ is obtained from the
PDF  which can be expressed as \cite{poor2013introduction}
\begin{equation}
f(\mathbf{y}|\boldsymbol{\theta})\propto\frac{1}{\left(\pi\sigma^{2}\right)^{PQ}}\exp\biggl({-\frac{1}{\sigma^{2}}\left(\mathbf{y}-\mathbf{m}\right)^{\mathsf{H}}\left(\mathbf{y}-\mathbf{m}\right)}\biggr),\label{log likelihood}
\end{equation}
and the log-likelihood function can be expressed as 
\begin{eqnarray}
L(\mathbf{y}) &{}={}&\ln{f(\mathbf{y}|\boldsymbol{\theta})}\nonumber\\
&{}={}&  -PQ\ln\pi\sigma^{2}-\frac{1}{\sigma^{2}}\left(\mathbf{y}-\mathbf{m}\right)^{\mathsf{H}}\left(\mathbf{y}-\mathbf{m}\right).\label{eq:loglikelihood}
\end{eqnarray}

\subsection{Cram\'{e}r-Rao Bound}

In this section, we derive the CRB of the channel parameters. 
Let $\tilde{\boldsymbol{\theta}}$
be the unbiased estimator of $\boldsymbol{\theta}$, the mean squared
error (MSE) is bounded by  \cite{trove.nla.gov.au/work/9004667} 
\begin{equation}
\mathbb{E}\left[\left(\tilde{\boldsymbol{\theta}}-\boldsymbol{\theta}\right)\left(\tilde{\boldsymbol{\theta}}-\boldsymbol{\theta}\right)^{\mathsf{T}}\right]\succeq\mathbf{J}_{\boldsymbol{\theta}}^{-1},
\end{equation}
where  $\mathbf{J}_{\boldsymbol{\theta}}$ is the Fisher information
matrix (FIM) defined as
\begin{equation}
\mathbf{J}_{\boldsymbol{\theta}}\triangleq-\mathbb{E}_{\mathbf{y|\boldsymbol{\theta}}}\left[\frac{\partial^{2}L(\mathbf{y})}{\partial\boldsymbol{\theta}\partial\boldsymbol{\theta}^{\mathsf{T}}}\right].\label{eq:jacobian}
\end{equation}
The FIM can be structured as 
\begin{eqnarray}
\mathbf{J}_{\boldsymbol{\theta}}  =  \left[\begin{array}{ccc}
\mathbf{J}(\phi_{m},\phi_{m}) & \quad\mathbf{J}(\phi_{m},\theta_{m}) & \quad \mathbf{J}(\phi_{m},\alpha_{m})\\
\mathbf{J}(\theta_{m},\phi_{m}) & \quad \mathbf{J}(\theta_{m},\theta_{m}) &\quad \mathbf{J}(\theta_{m},\alpha_{m})\\
\mathbf{J}(\alpha_{m},\phi_{m}) & \quad \mathbf{J}(\alpha_{m},\theta_{m}) & \quad \mathbf{J}(\alpha_{m},\alpha_{m})
\end{array}\right],\label{eq:jacob}
\end{eqnarray}
in which $\mathbf{J}(x_{t},x_{r})$
is defined as 
\begin{eqnarray} 
\mathbf{J}(x_{t},x_{r}) \triangleq - \mathbb{E}_{\mathbf{y|\boldsymbol{\theta}}}\left[\frac{\partial^{2}L(\mathbf{y})}{\partial x_{t}\partial x_{r}}\right].\label{eq:likelihood}
\end{eqnarray}
To evaluate the parameters of $\mathbf{J}_{\boldsymbol{\theta}}$, 
we consider the following Lemma 1. 

\begin{lem}
The trace and Kronecker product operations have the
following respective properties \cite{brewer1978kronecker,graham2018kronecker}
\begin{eqnarray}
&\mathbf{x}^{\mathsf{H}}\mathbf{A}\mathbf{y}  =\tr\left(\mathbf{A}\mathbf{x}^{\mathsf{H}}\mathbf{y}\right)=\tr\left(\mathbf{A}\mathbf{y}\mathbf{x}^{\mathsf{H}}\right), \\
&\left(\mathbf{A}\otimes\mathbf{B}\right)\left(\mathbf{C}\otimes\mathbf{D}\right)  =\left(\mathbf{A}\mathbf{C}\right)\otimes\left(\mathbf{B}\mathbf{D}\right). \label{eq:lemma1}
\end{eqnarray} 
\end{lem}

The entries of the matrix $\mathbf{J}_{\boldsymbol{\theta}}$ are given as follows 
\setlength{\arraycolsep}{0.0em}
	\begin{align}
	&\mathbf{J}(\phi_{m},\phi_{m})  =  \frac{2\alpha_{m}^2\pi^2E_s}{\sigma^{2}}\tr\biggl( \mathbf{A^{\mathsf{H}}}\mathbf{A}\big(\sin^{2}\phi_{m}\tilde{\mathbf{a}}_t(\phi_{m})\tilde{\mathbf{a}}_t^{\mathsf{H}}(\phi_{m})\big)
 \otimes\big(\mathbf{a}_r(\theta_{m})\mathbf{a}_r^{\mathsf{H}}(\theta_{m})\big)\biggr),\label{eq:firstenteries}\\
&\mathbf{J}(\phi_{m},\theta_{m}) = \frac{-2\alpha_{m}^2\pi^{2}E_s}{\sigma^{2}}\operatorname{Re}\biggl\{ \tr\biggl(\mathbf{A^{\mathsf{H}}}\mathbf{A}\big(\sin\phi_{m}\tilde{\mathbf{a}}_t(\phi_{m})\mathbf{a}_r^{\mathsf{T}}(\theta_{m})\big)\otimes\big(\sin\theta_{m}\mathbf{a}_t(\phi_{m})\tilde{\mathbf{a}}_r^{\mathsf{T}}(\theta_{m})\big)\biggr)\biggr\}, \\
&\mathbf{J}(\phi_{m},\alpha_{m})  = \frac{-2\alpha_{m}\pi E_s}{\sigma^{2}}\operatorname{Im}\biggl\{ \tr\biggl(\mathbf{A}^{\mathsf{H}}\mathbf{A}\big(\sin\phi_{m}\mathbf{a}_t^{*}(\phi_{m})\tilde{\mathbf{a}}_t^{\mathsf{H}}(\phi_{m})\big)\otimes\big(\mathbf{a}_r(\theta_{m})\mathbf{a}_r^{\mathsf{H}}(\theta_{m})\big)\biggr)\biggr\}, \\
&\mathbf{J}(\theta_{m},\theta_{m}) = \frac{2\alpha_{m}^2\pi^2E_s}{\sigma^{2}}\tr\biggl(\mathbf{A}^{\mathsf{H}}\mathbf{A}\big(\mathbf{a}_t^{*}(\phi_{m})\mathbf{a}_t^{\mathsf{T}}(\phi_{m})\big)
  \otimes\big(\tilde{\mathbf{a}}^*_{r}(\theta_{m})\tilde{\mathbf{a}}_r^{\mathsf{T}}(\theta_{m})\sin^{2}\theta_{m}\big)\biggr),\\
&\mathbf{J}(\theta_{m},\alpha_{m}) =\frac{2\alpha_{m}\pi E_s}{\sigma^{2}}\operatorname{Im}\biggl\{ \tr\biggl(\mathbf{A}^{\mathsf{H}}\mathbf{A}\big(\mathbf{a}_t^{*}(\phi_{m})\mathbf{a}_t^{\mathsf{T}}(\phi_{m})\big)
  \otimes\big(\sin\theta_{m}\mathbf{a}_r(\theta_{m})\tilde{\mathbf{a}}_r^{\mathsf{T}}(\theta_{m})\big)\biggr)\biggr\}, \\
&\mathbf{J}(\alpha_{m},\alpha_{m}) = \frac{2E_s}{\sigma^{2}}\tr\biggl(\mathbf{A}^{\mathsf{H}}\mathbf{A}\big(\mathbf{a}_t^{*}(\phi_{m})\mathbf{a}_t^{\mathsf{T}}(\phi_{m})\big)
   \otimes\big(\mathbf{a}_r(\theta_{m})\mathbf{a}_r^{\mathsf{H}}(\theta_{m})\big)\biggr). \label{eq: entries}
	\end{align}
The detail derivation of the entries is relegated to Appendix \ref{Proof of the entries}. 

\begin{rem}
As the essence of beam alignment is to select the beam pair
that meets a target rate, when the location information is accurately
known at the BS and the UE, the results converge to the optimal beam
with the actual values of $\boldsymbol{\theta}$ to realize a lossless
transmission. 
\end{rem}

\subsection{Downlink Channel Estimate and Rate Evaluation}
From the FIM, we express the  covariance matrix  of the estimation error of $\boldsymbol{\theta}$  as $\boldsymbol{\Sigma}_{\boldsymbol{\theta}}=\mathbf{J}_{\boldsymbol{\theta}}^{-1}$.
The error covariance matrix of $\mathbf{H}$ can be obtained by a Jacobian  transformation 
 matrix $\mathbf{T}$ given as 
 
\begin{equation}
\setcounter{equation}{54}
\boldsymbol{\Sigma}_{\mathbf{H}}=\mathbf{T}\boldsymbol{\Sigma}_{\boldsymbol{\theta}}\mathbf{T}^{\mathsf{T}},
\end{equation}
where 
	$\mathbf{T}={\partial\mathbf{h}}/{\partial\boldsymbol{\theta}}$,
 $\mathbf{h}=\operatorname{vec}(\mathbf{H})$ is obtained by the column vectorization of  the channel matrix $\mathbf{H}$. The  vector $\mathbf{h}$ can be expressed as 
\begin{eqnarray}
    \mathbf{h}&{}={}&\operatorname{vec}\left(\mathbf{H}\right)\nonumber\\
&{}={}&\sum_{m=0}^{M-1}\alpha_m\left(\mathbf{a}_t^{*}(\phi_{m})\otimes\mathbf{a}_r^{\mathsf{H}}(\theta_{m})\right).
\end{eqnarray}

Consequently, we obtain the entries of $\mathbf{T}$ as 
\begin{equation}
 \mathbf{T}=\left[
  \frac{\partial\mathbf{h}}{\partial\phi_m},  \frac{\partial\mathbf{h}}{\partial\theta_m},  \frac{\partial\mathbf{h}}{\partial\alpha_m}
 \right],
\end{equation}
where
\setlength{\arraycolsep}{0.0em}
\begin{eqnarray}
	\frac{\partial\mathbf{h}}{\partial\phi_m}&{}={}&-\alpha j\pi \sin\phi_m\left(\tilde{\mathbf{a}}_t(\phi_{m})\otimes\mathbf{a}_r(\theta_{m})\right), \\
	\frac{\partial\mathbf{h}}{\partial\theta_m}&{}={}&\alpha j\pi \sin\theta_m\left(\mathbf{a}_t^{*}(\phi_{m})\otimes\tilde{\mathbf{a}}^{*}_{r}(\theta_{m})\right), \\
	\frac{\partial\mathbf{h}}{\partial\alpha_m}&{}={}& \left(\mathbf{a}_t^{*}(\phi_{m})\otimes\mathbf{a}_r(\theta_{m})\right).
\end{eqnarray}
The rate can then be computed with the estimated channel  $\hat{\mathbf{H}}=\mathbf{H}+\mathbf{E}$ as follows 
\setlength{\arraycolsep}{0.0em}
\begin{align}
 R_d(\mathbf{u}_{\text{op}},\mathbf{v}_{\text{op}}) 
 &= \log_{2}\left(1+\frac{E_s\left|\mathbf{u}_{\text{op}}^{\mathsf{H}}\mathbf{\hat{H}}\mathbf{v}_{\text{op}}\right|^{2}}{\sigma^2+E_s\mathbb{E}\left\{ \left|\mathbf{u}_{\text{op}}^{\mathsf{H}}\left(\mathbf{\hat{H}}-\mathbf{H}\right)\mathbf{v}_{\text{op}}\right|^{2}\right\} }\right)\nonumber\\
& = \log_{2}\left(1+\frac{E_s\left|\mathbf{u}_{\text{op}}^{\mathsf{H}}\mathbf{\hat{H}}\mathbf{v}_{\text{op}}\right|^{2}}{\sigma^2+E_s \mathbb{E}\left\{ \left|\mathbf{u}_{\text{op}}^{\mathsf{H}}\mathbf{E}\mathbf{v}_{\text{op}}\right|^{2}\right\} }\right),\label{eq:rateestimate}
\end{align}
 where $\mathbf{E}\sim\mathcal{CN}(\mathbf{0},\boldsymbol{\Sigma}_{\mathbf{H}}^2)$. 

\section{Numerical Results}

In this section, we present the simulation setup and show the performance of the proposed beam alignment algorithm.

\subsection{Simulation Setup}

 We consider a scenario with two potential reflecting points as shown in Fig.
\ref{fig:Scenario}. 
The actual location coordinates of the BS, two reflecting points and the UE are given as $\mathbf{l}_{\text{BS}}=[0,0]$, $\mathbf{l}_{1}=[50,50]$, $\mathbf{l}_{2}=[50,-50]$ and $\mathbf{l}_{\text{UE}}=[100,0]$ respectively.
Furthermore, the maximum location error of the BS, two reflecting points and the UE  when observed from the BS are given by $r_{\text{BS}}^{\text{BS}}=0$, $r_{1}^{\text{BS}}=11$, $r_{2}^{\text{BS}}=15$ and $r_{\text{UE}}^{\text{BS}}=13$ respectively, while the  maximum location error  of the BS and the reflecting points  when observed from the UE are given as $r_{\text{BS}}^{\text{UE}}=0$, $r_{1}^{\text{UE}}=18$, $r_{2}^{\text{UE}}=17$ and $r_{\text{UE}}^{\text{UE}}=7$ respectively, where  all measurements are in meters.
 Furthermore, the  maximum location estimation errors  are only used to validate the proposed algorithm and also allow us to compare the proposed beam alignment scheme with the two-step beam alignment in \cite{maschietti2017robust}.
 
\begin{figure}
	\centering{}
	\includegraphics[width=0.6\columnwidth]{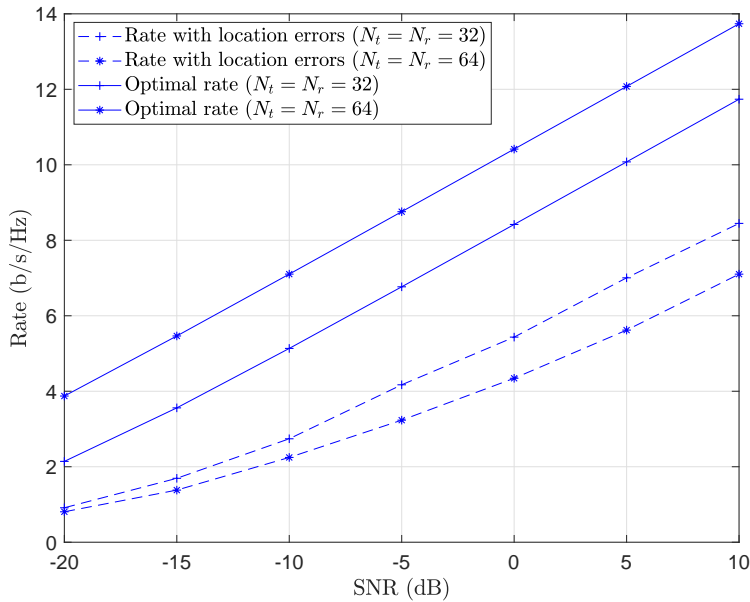}\caption{Rate vs SNR for varying number of antennas in the uncertainty region. \label{fig:Effect-of-location}}
\end{figure}

As discussed in Section II, we focus on the initial access phase,
and the simulations are performed over $10^3$ independent blocks. The location estimation and beam alignment are
performed at the beginning of each block as summarized in Algorithm
I. The metric used to select a beam is the rate loss of the beam compared with the optimal beam pair, where the optimal beams pair are achieved with the precise location information and perfect channel state information.
We set the target rate to $95\%$ of the optimal rate, i.e., $R_0=0.95 \max_{\mathbf{u},\mathbf{v}}R_d(\mathbf{u},\mathbf{v})$ except stated otherwise, the size of the window is set to the number of beams that can be processed in each slot (i.e., $W_{\text{BS}}=N_b$). 
\subsection{Results and Discussion}

Fig. \ref{fig:Effect-of-location} shows the rate  as a function of the SNR obtained from the beam vectors corresponding to the initial estimate of $\hat{\phi}_m$ and $\hat{\theta}_m$ in the uncertainty region.  The results are shown with varying number of antennas. A performance loss can be observed when compared with the rates obtained from selecting the optimal beams.  From  (\ref{eq:codebookrx}) and (\ref{eq:codebooktx}), as the number of antennas increases, the beamwidth becomes narrow, and the severity of beam  misalignment increases. As observed from the figure, the use of a low number of array antenna (lower array gain) may outperform a high number of antenna array depending on the severity of the location estimation error. In addition,  the result shows that although location information with some degree of precision is expected to be available to 5G systems, beam alignment  is required in the uncertainty region of the location information especially for a large number of antenna arrays with narrow beamwidth. 

\begin{figure}
	\centering{}\includegraphics[width=0.6\columnwidth]{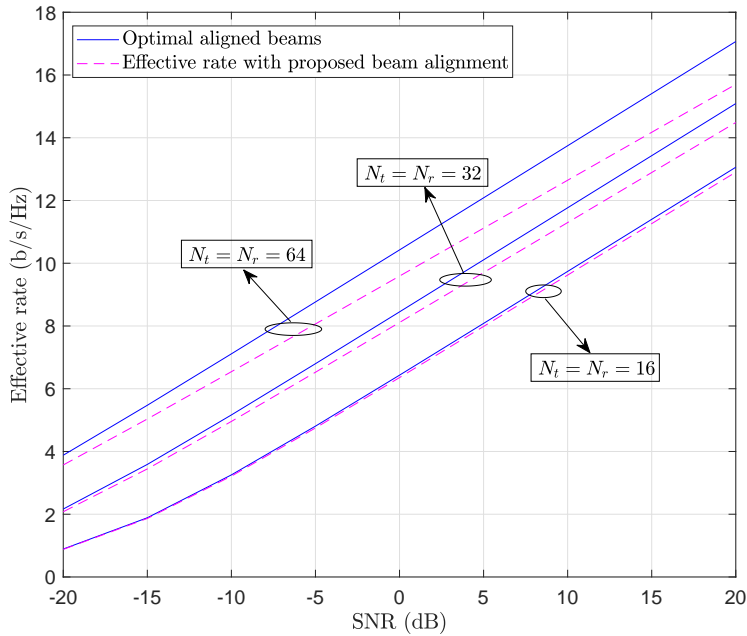}\caption{Comparison of rate vs SNR, with varying number of antennas. Fixed parameters: $N_b=5,N=100$.\label{fig:Rate-vs-SNR,}} 
\end{figure}
In Fig. \ref{fig:Rate-vs-SNR,}, we compare the proposed beam alignment scheme with the optimal beam selection scheme. 
When Fig. \ref{fig:Rate-vs-SNR,} is compared with Fig. \ref{fig:Effect-of-location},
the effectiveness of the proposed beam alignment can be observed from
the rates achieved. When the proposed beam alignment is compared with
the optimal aligned beams plots, a performance gap is observed which is due to the  beam alignment overhead given by  (\ref{eq:penalty}).
However, to achieve the optimal beam pair, we assume that the channel information of each node is perfectly known  which is difficult to achieve in practice. Furthermore, the result shows that exploiting location  information can further reduce the beam alignment overhead with a slight  penalty on the performance. The beam alignment overhead can be reduced with high precision in the location information. In addition, it can be observed that the gap between the proposed beam alignment scheme and the optimal aligned beam plots increases with an increasing number of antenna which is mainly due to the fact that as the number of antenna increases, the beamwidth decreases. Hence, as the number of beams in the uncertainty region increases, the number of slots $N_a$ required for beam alignment also increases. 

The  proposed algorithm is compared with the results obtained from the Two-Step algorithm in  \cite{maschietti2017robust},
with $N_{t}=N_{r}=64$ antennas in Fig \ref{fig:Rate-vs-SNR,-1}. From Fig \ref{fig:Rate-vs-SNR,-1}, it can be observed that the proposed scheme outperforms the two step  algorithm proposed in \cite{maschietti2017robust}, even though the penalty of beam alignment is considered in our proposed scheme. This is because in the proposed algorithm, the beam alignment is jointly coordinated by the BS and UE, while in \cite{maschietti2017robust}, the beam alignment is carried out independently at each node. 

\begin{figure}
	\centering{}\includegraphics[width=0.6\columnwidth]{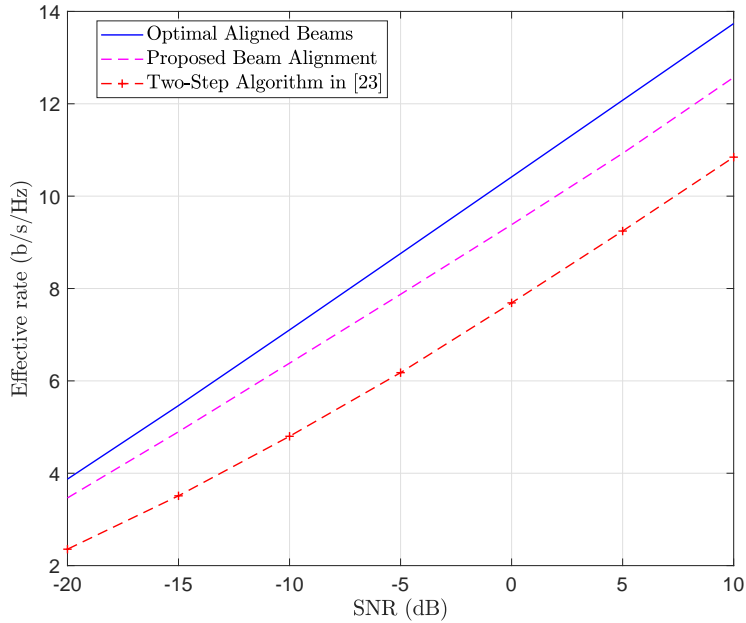}\caption{Comparison of proposed scheme  and  the two step algorithm in \cite{maschietti2017robust}. Fixed parameters: $N_{t}=N_{r}=64,N_b=5,N=100$.
		\label{fig:Rate-vs-SNR,-1} }
\end{figure} 

\begin{figure}
	\centering{}\includegraphics[width=0.6\columnwidth]{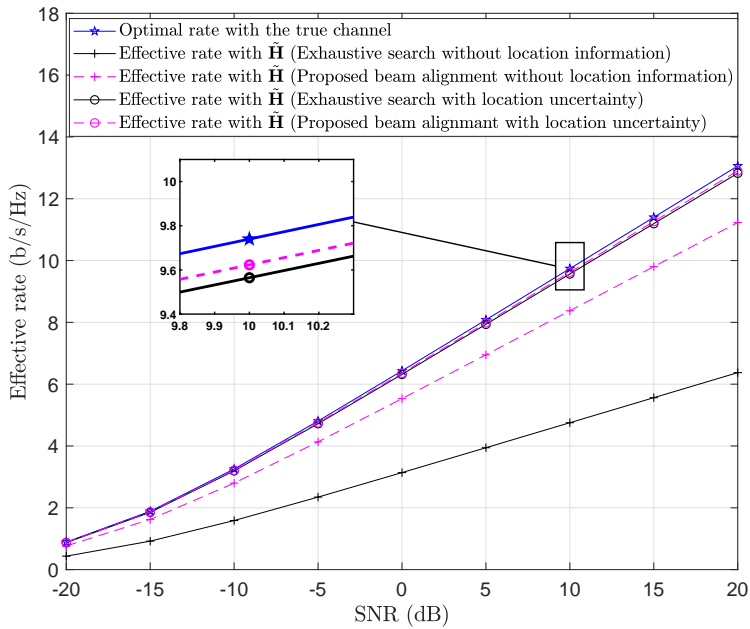}\caption{Effective rates with channel estimate as a function of SNR. Fixed parameters: $N_t= N_r=16,N_b=5,N=100$, $R_0=\max_{\mathbf{u},\mathbf{v}}R_d(\mathbf{u},\:\mathbf{v})$. \label{fig:CRB} }
\end{figure} 

Fig. \ref{fig:CRB} compares the effective rates obtained with the channel estimate as a function of SNR.  The plots without location information are obtained by transmitting with the beam set  ($\mathcal{V}$ and $\mathcal{U}$). With the knowledge of the location uncertainty region, beam alignment is performed using the beam set  ($\mathcal{B}_{\text{BS}}^m$ and $\mathcal{B}_{\text{UE}}^m$),
where $m$ correspond to the LOS path in this plot. In this result, we evaluate the FIM given by (\ref{eq:jacob}) and the rates are computed  from (\ref{eq:rateestimate}), where the expectation is taken over the channel estimation error matrix. In the plot of the exhaustive search without location information, the channel is estimated by transmitting  symbols with each of the 16-dimensional beams vectors in $\mathcal{V}$ and $\mathcal{U}$ since $N_{t}=N_{r}=16$. A performance gap can be observed when the proposed beam alignment is compared with the exhaustive search without location information. The improvement in the proposed algorithm is due to the fast beam alignment procedure discussed in Section \ref{beam alignement}, leading to a small beam alignment penalty when compared to the exhaustive search.  
With the location information and maximum location estimation error known \textit{a priori}, the channel parameters can be estimated by transmitting with the few beams in  $\mathcal{B}_{\text{BS}}^m$ and $\mathcal{B}_{\text{UE}}^m$ which correspond to the  set of beams vectors in the uncertainty region of the $m$-th path. We plot the effective rate  of the exhaustive search and the proposed beam alignment algorithm in the location uncertainty region. It is observed that for a fixed $N_b$ and $N$, the  performance of both schemes is quite close to the optimal rate plot. This is because there are few beams in the location uncertainty region and hence, the beam alignment overhead is low.

\begin{figure}
	\centering{}\includegraphics[width=0.6\columnwidth]{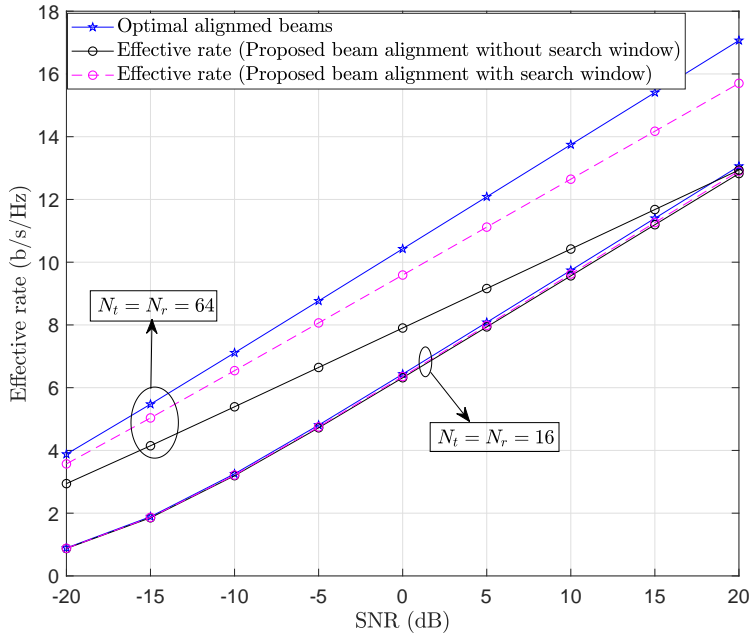}\caption{Comparison of the proposed beam alignment scheme with and without search window in the uncertainty region. Fixed parameters: $N_b=5$, $N=100$, $R_0= \max_{\mathbf{u},\mathbf{v}}R_d(\mathbf{u},\:\mathbf{v})$ . \label{fig:CRBcomparison} }
\end{figure}

As the number of beams in the uncertainty region increases due to narrow beams at the BS and UE (increasing $N_r$ and $N_t$), the effect of the search window can be observed, and the efficiency of the proposed scheme with search window  become more noticeable  as shown in Fig. \ref{fig:CRBcomparison}.
In the figure, the effective rate of the proposed scheme  is plotted showing the effect of the proposed window  for varying number of antennas. As the number of antennas increase from $N_t=N_r=16$ to $N_t=N_r=64$, the number of beams in the uncertainty region also increase, thereby requiring more slots for beam alignment. However, with the implementation of the search window, the proposed scheme is able to achieve fast beam alignment  compared to the plot of the proposed beam alignment scheme without the implementation of the search window, hence, a better effective rate performance is achieved as shown in the result. The improvement  obtained from the use of the search  window in the proposed scheme is as a result the reduction in the number of beams to be searched after determining the direction of search in the window.

From Fig. \ref{fig:CRB} and Fig. \ref{fig:CRBcomparison}, it can be deduced that with high precision in location information, there is no need to estimate the channel with a large number of beam vectors from $\mathcal{V}$ and $\mathcal{U}$  since similar performance can be achieved with much fewer beam set in the uncertainty region.   Intuitively, with some degree of precision of the location information at the BS and UE and  the knowledge of the maximum location error, the number of beams required to achieve the best estimate of $\theta$ and $\phi$ can be reduced. Hence,  exploiting location  information in the proposed scheme reduces the beam alignment overhead.    

\section{Conclusion}

In this paper, we propose a coordinated beam alignment algorithm which
exploits the noisy location information of  the UE, and potential reflecting points.
The proposed beam alignment enables the UE and BS to jointly coordinate
the beam alignment to combat the path-loss in the mmWave range. The scheme
speeds up the beam alignment procedure by focusing the search in a
specific region bounded by the location error margin. Furthermore,
an intelligent search is performed in this specific region by searching
across a small window of beams to determine the direction of the actual
beam. The numerical results show that the proposed algorithm can significantly
improve the beam alignment speed when compared to the scenario with
perfect channel information and other existing schemes.

\appendices
\section{Proof of Proposition I\label{proof proposition 1}}
\begin{IEEEproof} Without loss of generality, we define $\epsilon_m={\gamma}_m-{\hat{\phi}}_{m}$ as the half angle of the uncertainty region in path $m$ from the BS as shown in Fig. \ref{fig:Error-boundary}, where ${\gamma}_m$ is the angle  between the antenna normal and the maximum location error boundary  of the node in path $m$. Then we can express
	\begin{align} 
	y_{m} & =\hat{d}_{m}^{\text{BS}}\sin{\hat{\phi}}_{m},\label{eq:x and y}\\
	x_{m} & =\hat{d}_{m}^{\text{BS}}\cos{\hat{\phi}}_{m},  
	\end{align} 
	where $\hat{d}_{m}^{\text{BS}}$ can be obtained from equation (\ref{eq:BSdistance}) for the BS,
	and (\ref{eq:UEdistance}) for the UE. From simple geometry, we can obtain 
	\begin{align} 
	\tan{\gamma}_m & =\frac{y_{m}+r_m^{\text{BS}}}{x_{m}}\nonumber \\
	& =\frac{\hat{d}_{m}^{\text{BS}}\sin{\hat{\phi}}_{m}+r_m^{\text{BS}}}{\hat{d}_{m}^{\text{BS}}\cos{\hat{\phi}}_{m}}, \quad m=1,\ldots,M,
	\end{align} by solving  for ${\gamma}_m$, the following can be obtained 
	\begin{equation}
	{\gamma}_m=\arctan\left(\frac{\hat{d}_{m}^{\text{BS}}\sin{\hat{\phi}}_{m}+r_m^{\text{BS}}}{\hat{d}_{m}^{\text{BS}}\cos{\hat{\phi}}_{m}}\right)\label{eq:phiang}.
	\end{equation}
	By solving for
	$\epsilon_m={\gamma}_m-{\hat{\phi}}_{m}$, equation (\ref{eq:phiang}) can be expressed as
	 \begin{equation}
	\epsilon_m+{\hat{\phi}}_{m}=\arctan\left(\frac{\hat{d}_{m}^{\text{BS}}\sin{\hat{\phi}}_{m}+r_m^{\text{BS}}}{\hat{d}_{m}^{\text{BS}}\cos{\hat{\phi}}_{m}}\right). \label{eq:epsilon}
	\end{equation}
	Following similar steps from (\ref{eq:x and y}) to (\ref{eq:epsilon}) we obtain
	\begin{equation}
	\delta_m+{\hat{\theta}}_{m}=\arctan\left(\frac{\hat{d}_{m}^{\text{UE}}\sin{\hat{\theta}}_{m}+r_m^{\text{UE}}}{\hat{d}_{m}^{\text{UE}}\cos{\hat{\theta}}_{m}}\right),
	\end{equation}
	which concludes the proof.
	
\end{IEEEproof} 


\section{Proof of the entries of the Matrix $\mathbf{J}_{\boldsymbol{\theta}}$\label{Proof of the entries}}

In this section, we present the derivation of the entries of the $\mathbf{J}_{\boldsymbol{\theta}}$. 

\begin{IEEEproof} The proof begins with the  calculations of
\begin{equation}
    \mathbf{J}(\phi_{m},\phi_{m})\triangleq-\mathbb{E}_{\mathbf{y|\boldsymbol{\theta}}}\left[\frac{\partial^{2} L(\mathbf{y})}{\partial\phi_{m}^{2}}\right].
\end{equation}
Note that $L(\mathbf{y})$
	in (\ref{eq:likelihood}) is given by the log-likelihood function
	in (\ref{eq:loglikelihood}) and by differentiating $L(\mathbf{y})$
	w.r.t $\phi_{m}$, we obtain
	\begin{eqnarray}
	\frac{\partial L(\mathbf{y})}{\partial\phi_{m}} = \frac{E_s}{\sigma^{2}}\left[\left(\mathbf{y}-\mathbf{m}\right)^{\mathsf{H}}\frac{\partial\mathbf{m}}{\partial\phi_{m}}+\frac{\partial\mathbf{m}^{\mathsf{H}}}{\partial\phi_{m}}\left(\mathbf{y}-\mathbf{m}\right)\right],\label{first dericative}
	\end{eqnarray}
	taking the second differential of (\ref{first dericative}) w.r.t $\phi_{m}$
	we have 
	\setlength{\arraycolsep}{0.0em}
	\begin{eqnarray}
	\frac{\partial^{2}L(\mathbf{y})}{\partial\phi_{m}^{2}} &{}={}& \frac{E_s}{\sigma^{2}}\left[-\frac{\partial\mathbf{m}^{\mathsf{H}}}{\partial\phi_{m}}\frac{\partial\mathbf{m}}{\partial\phi_{m}}+\left(\mathbf{y}-\mathbf{m}\right)^{\mathsf{H}}\frac{\partial^{2}\mathbf{m}}{\partial\phi_{m}^{2}}\right.\label{eq:second dericative}{+} \left.\frac{\partial^{2}\mathbf{m}^{\mathsf{H}}}{\partial\phi_{m}^{2}}\left(\mathbf{y}-\mathbf{m}\right)-\frac{\partial\mathbf{m}^{\mathsf{H}}}{\partial\phi_{m}}\frac{\partial\mathbf{m}}{\partial\phi_{m}}\right].
	\end{eqnarray}
	By taking the expectation of (\ref{eq:second dericative}) w.r.t.
	$L(\mathbf{y})$ we obtain 
	\begin{eqnarray}
	\mathbf{J}(\phi_{m},\phi_{m})  =  \frac{2E_s}{\sigma^{2}}\left[\frac{\partial\mathbf{m}^{\mathsf{H}}}{\partial\phi_{m}}\frac{\partial\mathbf{m}}{\partial\phi_{m}}\right].\label{eq:jacodtxsq}
	\end{eqnarray}
	Next, we will determine $\partial\mathbf{m}^{\mathsf{H}}/{\partial\phi_{m}}$,
	where $\mathbf{m}$ is given by (\ref{eq:mean}) 
	\setlength{\arraycolsep}{0.0em}
	\begin{eqnarray}
	\frac{\partial\mathbf{m}^{\mathsf{H}}}{\partial\phi_{m}} &{}={}&\frac{\partial}{\partial\phi_{m}}\alpha_{m}\biggl(\mathbf{a}_t^{\mathsf{T}}(\phi_{m})\otimes\mathbf{a}_r^{\mathsf{H}}(\theta_{m})\biggr)\mathbf{A^{\mathsf{H}}}\nonumber \\
	&{}={}&\alpha_{m}j\pi\sin\phi_{m}\biggl(\tilde{\mathbf{a}}_t^{\mathsf{H}}(\phi_{m})\otimes\mathbf{a}_r^{\mathsf{H}}(\theta_{m})\biggr)\mathbf{A^{\mathsf{H}}},\label{eq:dmdtxprime}
	\end{eqnarray}
	where 
\setlength{\arraycolsep}{0.0em}
	\begin{equation}
	    \begin{split}
	\tilde{\mathbf{a}}_t(\phi_{m})=\frac{1}{\sqrt{N_{t}}}\biggl[0,\;e^{-j\pi\cos\phi_{m}},\ldots,(N_{t}-1)e^{-j\pi(N_{t}-1)\cos\phi_{m}}\biggr]^{\mathsf{T}}.
	\end{split}
	\end{equation}
	
	 Similarly, $\partial\mathbf{m}/{\partial\phi_{m}}$ can
	be obtained as 
	\begin{eqnarray}
	\frac{\partial\mathbf{m}}{\partial\phi_{m}}  =  -\mathbf{A}\alpha_{m}j\pi\sin\phi_{m}\bigg(\tilde{\mathbf{a}}_t(\phi_{m})\otimes\mathbf{a}_r(\theta_{m})\bigg).\label{eq:dmdtx}
	\end{eqnarray}
	By substituting (\ref{eq:dmdtxprime}) and (\ref{eq:dmdtx}) in (\ref{eq:jacodtxsq}),
	we obtain 
	\setlength{\arraycolsep}{0.0em}
\setlength{\arraycolsep}{0.0em}
\begin{align}
\begin{split}
\mathbf{J}(\phi_{m},\phi_{m}) =&\frac{2\alpha_{m}^2\pi^2E_s}{\sigma^{2}}\tr\biggl(\mathbf{A^{\mathsf{H}}}\mathbf{A}\big(\sin^{2}\phi_{m}\tilde{\mathbf{a}}_t(\phi_{m})\tilde{\mathbf{a}}_t^{\mathsf{H}}(\phi_{m})\big)
{\otimes}\big(\mathbf{a}_r(\theta_{m})\mathbf{a}_r^{\mathsf{H}}(\theta_{m})\big)\biggr),
\end{split}
\end{align}
where 
	 \setlength{\arraycolsep}{0.0em}
\begin{equation}
	    \begin{split}
	\tilde{\mathbf{a}}_r(\theta_{m})
	=\frac{1}{\sqrt{N_{r}}}\biggl[0,e^{-j\pi\cos \theta_{m}},\ldots,\biggl.(N_{r}-1)e^{-j\pi(N_{r}-1)\cos \theta_{m}}\biggr]^{\mathsf{T}}.
	\end{split}
	\end{equation}
	
	Next, we compute
	\begin{equation}
	  \mathbf{J}(\phi_{m},\theta_{m})\triangleq-\mathbb{E}_{\mathbf{y|\boldsymbol{\theta}}}\left[\frac{\partial^{2}L(\mathbf{y}) }{\partial\phi_{m}\partial\theta_{m}}\right],\label{eq:jphitheta}
	\end{equation}
	where we take the second derivative of (\ref{first dericative}) with
	respect to $\phi_{m}$ and $\theta_{m}$ given as 
	\begin{eqnarray}
	\frac{\partial^{2}L(\mathbf{y})}{\partial\phi_{m}\partial\theta_{m}}  = \frac{E_s}{\sigma^{2}}\left[-\frac{\partial\mathbf{m}^{\mathsf{H}}}{\partial\phi_{m}}\frac{\partial\mathbf{m}}{\partial\theta_{m}}+\frac{\partial\mathbf{m}^{\mathsf{H}}}{\partial\theta_{m}}\frac{\partial\mathbf{m}}{\partial\phi_{m}}\right].\label{eq:dsqfxdtdr}
	\end{eqnarray}
	The partial derivative of $\mathbf{m}$ w.r.t $\theta_{m}$ can
	be obtained as 
	\begin{eqnarray}
	\frac{\partial\mathbf{m}}{\partial\theta_{m}}  =  \alpha_{m}j\pi\sin\theta_{m}\mathbf{A}\big(\mathbf{a}_t^{*}(\phi_{m})\otimes\tilde{\mathbf{a}}_r^{*}(\theta_{m})\big),\label{eq:dmdr}
	\end{eqnarray}
	where the partial derivative of $\mathbf{m}^{\mathsf{H}}$ w.r.t $\theta_{m}$
	can be obtained similarly as 
	\begin{equation}
	\frac{\partial\mathbf{m}^{\mathsf{H}}}{\partial\theta_{m}}  =  -\alpha_{m}j\pi\sin\theta_{m}\big(\mathbf{a}_t^{\mathsf{T}}(\phi_{m})\otimes\tilde{\mathbf{a}}_r^{\mathsf{T}}(\theta_{m})\big)\mathbf{A}^{\mathsf{H}}.\label{eq:dmprimdr}
	\end{equation}
	By substituting (\ref{eq:dmdtxprime}), (\ref{eq:dmdtx}), (\ref{eq:dmdr})
	and (\ref{eq:dmprimdr}) into (\ref{eq:dsqfxdtdr}), 
	\setlength{\arraycolsep}{0.0em}
and using the properties in Lemma 1, (\ref{eq:jphitheta}) can be simplified
	as 
	\setlength{\arraycolsep}{0.0em}
%
	\begin{align}
	        \mathbf{J}\left(\phi_{m},\theta_{m}\right)
	        =\frac{-2\alpha_{m}^2\pi^2E_s}{\sigma^{2}}\operatorname{Re}\biggl\{\tr\biggl(\mathbf{A^{\mathsf{H}}}\mathbf{A}\big(\sin\phi_{m}\tilde{\mathbf{a}}_t(\phi_{m})\mathbf{a}_r^{\mathsf{T}}(\theta_{m})\big)
	        {\otimes}\big(\sin\theta_{m}\mathbf{a}_t(\phi_{m})\tilde{\mathbf{a}}_r^{\mathsf{T}}(\theta_{m})\big)\biggr)\biggr\},
	\end{align}
	where Re$\left\{ x\right\} $ denotes the real part of the complex
	variable $x$. 
	
	Next, we compute
	\begin{equation}
	    \mathbf{J}(\phi_{m},\alpha_{m})\triangleq-\mathbb{E}_{\mathbf{y|\boldsymbol{\theta}}}\left[\frac{\partial^{2}L(\mathbf{y}) }{\partial\phi_{m}\partial
	    \alpha_{m}}\right],
	\end{equation} 
	which can be obtained by taking the second derivative of (\ref{first dericative})
	with respect to $\phi_{m}$ and $\alpha_{m}$ given as, 
	\begin{eqnarray}
	\frac{\partial^{2}L(\mathbf{y})}{\partial\phi_{m}\partial\alpha_{m}}  = \frac{E_s}{\sigma^{2}}\left[\frac{\partial\mathbf{m}^{\mathsf{H}}}{\partial\phi_{m}}\frac{\partial\mathbf{m}}{\partial\alpha_{m}}+\frac{\partial\mathbf{m}^{\mathsf{H}}}{\partial\alpha_{m}}\frac{\partial\mathbf{m}}{\partial\phi_{m}}\right],\label{eq: dpathgain}
	\end{eqnarray}
	where 
	\begin{eqnarray}
	\frac{\partial\mathbf{m}}{\partial\alpha_{m}}  =  \mathbf{A}\big(\mathbf{a}_t^{*}(\phi_{m})\otimes\mathbf{a}_r(\theta_{m})\big),\label{eq:dmdalpa}
	\end{eqnarray}
	and 
	\begin{eqnarray}
	\frac{\partial\mathbf{m}^{\mathsf{H}}}{\partial\alpha_{m}}  =  \big(\mathbf{a}_t^{\mathsf{T}}(\phi_{m})\otimes\mathbf{a}_r^{\mathsf{H}}(\theta_{m})\big)\mathbf{A}^{\mathsf{H}}.\label{eq:dmprimesalpha}
	\end{eqnarray}
	By using the properties defined in Lemma 1 and substituting (\ref{eq:dmdtxprime}),
	(\ref{eq:dmdtx}), (\ref{eq:dmdalpa}) and (\ref{eq:dmprimesalpha})
	into (\ref{eq: dpathgain}) we obtain 
	\setlength{\arraycolsep}{0.0em}
	\begin{align}
	\mathbf{J}(\phi_{m},\alpha_{m}) 
	=\frac{-2\alpha_{m}\pi E_s}{\sigma^{2}}\operatorname{Im}\biggl\{ \tr\biggl(\mathbf{A}^{\mathsf{H}}\mathbf{A}\big(\sin\phi_{m}\mathbf{a}_t^{*}(\phi_{m})\tilde{\mathbf{a}}_t^{\mathsf{H}}(\phi_{m})\big)
	  \otimes\left(\mathbf{a}_r(\theta_{m})\mathbf{a}_r^{\mathsf{H}}(\theta_{m})\right)\biggr)\biggr\} ,
	\end{align}
	where Im$\left\{ x\right\} $ is the imaginary part of the complex
	variable $x$. 
	
	Next, we determine $\mathbf{J}(\theta_{m},\theta_{m})$ from
	\begin{equation}
	    \mathbf{J}(\theta_{m},\theta_{m})\triangleq-\mathbb{E}_{\mathbf{y|\boldsymbol{\theta}}}\left[\frac{\partial^{2}L(\mathbf{y}) }{\partial\theta_{m}^2}\right],
	\end{equation}
	similar
	to (\ref{eq:jacodtxsq}) as follows 
	\begin{eqnarray}
	\mathbf{J}(\theta_{m},\theta_{m})  =  \frac{2E_s}{\sigma^{2}}\left[\frac{\partial\mathbf{m}^{\mathsf{H}}}{\partial\theta_{m}}\frac{\partial\mathbf{m}}{\partial\theta_{m}}\right],\label{eq:jacobrxsq}
	\end{eqnarray}
	which can be evaluated by substituting (\ref{eq:dmdr}) and (\ref{eq:dmprimdr})
	into (\ref{eq:jacobrxsq}) as 
	\begin{eqnarray}
	\mathbf{J}(\theta_{m},\theta_{m}) &{} ={}& \frac{2\alpha_{m}^2\pi^2E_s\sin^{2}\theta_{m}}{\sigma^{2}}\biggl[\big(\mathbf{a}_t^{\mathsf{T}}(\phi_{m})\otimes\tilde{\mathbf{a}}_r^{\mathsf{T}}(\theta_{m})\big)
	 \times\mathbf{A}^{\mathsf{H}}\mathbf{A}\left(\mathbf{a}_t^{*}(\phi_{m})\otimes\tilde{\mathbf{a}}_r^{*}(\theta_{m})\right)\biggr] \nonumber\\
	  &{} ={}&  \frac{2\alpha_{m}^2\pi^2E_s}{\sigma^{2}}\tr\biggl(\mathbf{A}^{\mathsf{H}}\mathbf{A}\big(\mathbf{a}_t^{*}(\phi_{m})\mathbf{a}_t^{\mathsf{T}}(\phi_{m})\big)
	 \otimes\big(\tilde{\mathbf{a}}_r(\theta_{m})\tilde{\mathbf{a}}_r^{\mathsf{H}}(\theta_{m})\sin^{2}\theta_{m}\big)\biggr).
	\end{eqnarray}
	
	Next, we evaluate the entries of  $\mathbf{J}(\theta_{m},\alpha_{m})$ defined as 
		\begin{equation}
	    \mathbf{J}(\theta_{m},\alpha_{m})\triangleq-\mathbb{E}_{\mathbf{y|\boldsymbol{\theta}}}\left[\frac{\partial^{2}L(\mathbf{y}) }{\partial\theta_{m}\partial
	    \alpha_{m}}\right],
	\end{equation}
	where the second derivative of (\ref{first dericative}) is taken with respect to 
	respect to $\theta_{m}$ and $\alpha_{m}$ as follows
	
	\begin{eqnarray}
	\mathbf{J}(\theta_{m},\alpha_{m})  =  \frac{E_s}{\sigma^{2}}\left[\frac{\partial\mathbf{m}^{\mathsf{H}}}{\partial\theta_{m}}\frac{\partial\mathbf{m}}{\partial\alpha_{m}}+\frac{\partial\mathbf{m}^{\mathsf{H}}}{\partial\alpha_{m}}\frac{\partial\mathbf{m}}{\partial\theta_{m}}\right],\label{eq:jacobthetaalpha}
	\end{eqnarray}
	which can be evaluated by substituting (\ref{eq:dmdr}), (\ref{eq:dmprimdr}),
	(\ref{eq:dmdalpa}) and (\ref{eq:dmprimesalpha}) into (\ref{eq:jacobthetaalpha})
	as follows
	\begin{eqnarray}
	\mathbf{J}(\theta_{m},\alpha_{m}) &{} ={}&  \frac{2\alpha_{m}\pi E_s}{\sigma^{2}}\sin\theta_{m}\operatorname{Im}\biggl\{\big(\mathbf{a}_t^{\mathsf{T}}(\phi_{m})\otimes\tilde{\mathbf{a}}_r^{\mathsf{T}}(\theta_{m})\big)
	\times\mathbf{A}^{\mathsf{H}}\mathbf{A}\big(\mathbf{a}_t^{*}(\phi_{m})\otimes\mathbf{a}_r(\theta_{m})\big)\biggr\} \nonumber\\
	&{} ={}&  \frac{2\alpha_{m}j\pi E_s}{\sigma^{2}}\operatorname{Im}\biggl\{ \tr\biggl(\mathbf{A}^{\mathsf{H}}\mathbf{A}\big(\mathbf{a}_t^{*}(\phi_{m})\mathbf{a}_t^{\mathsf{T}}(\phi_{m})\big)
	 \otimes\big(\sin\theta_{m}\mathbf{a}_r(\theta_{m})\tilde{\mathbf{a}}_r^{\mathsf{T}}(\theta_{m})\big)\biggr)\biggr\}. 
	\end{eqnarray}
	
	Finally, the factor
	\begin{equation}
	    \mathbf{J}(\alpha_{m},\alpha_{m})\triangleq-\mathbb{E}_{\mathbf{y|\boldsymbol{\theta}}}\left[\frac{\partial^{2}L(\mathbf{y})}{\partial\alpha_{m}^2}\right]
	\end{equation}
	can be evaluated as: 
	\begin{eqnarray}
	\mathbf{J}(\alpha_{m},\alpha_{m})= \frac{2E_s}{\sigma^{2}}\left[\frac{\partial\mathbf{m}^{\mathsf{H}}}{\partial\alpha_{m}}\frac{\partial\mathbf{m}}{\partial\alpha_{m}}\right],\label{eq:jacobalphasq}
	\end{eqnarray}
	where the parameters $\partial\mathbf{m}^{\mathsf{H}}/{\partial\alpha_{m}}$
	and ${\partial\mathbf{m}}/{\partial\alpha_{m}}$ can be obtained
	by substituting (\ref{eq:dmdalpa}) and (\ref{eq:dmprimesalpha})
	into (\ref{eq:jacobalphasq}), from which we obtain 
	\setlength{\arraycolsep}{0.0em}
	\begin{align}
	    \mathbf{J}(\alpha_{m},\alpha_{m}) 
	&= \frac{2E_s}{\sigma^{2}}\tr\biggl(\big(\mathbf{a}_t^{\mathsf{T}}(\phi_{m})\otimes\mathbf{a}_r^{\mathsf{H}}(\theta_{m})\mathbf{A}^{\mathsf{H}}\mathbf{A}\big(\mathbf{a}_t^{*}(\phi_{m})\otimes\mathbf{a}_r(\theta_{m})\big)\biggr)\nonumber\\
	&=\frac{2E_s}{\sigma^{2}}\tr\biggl(\mathbf{A}^{\mathsf{H}}\mathbf{A}\big(\mathbf{a}_t^{*}(\phi_{m})\otimes\mathbf{a}_r(\theta_{m})\big)  \big(\mathbf{a}_t^{\mathsf{T}}(\phi_{m})\otimes\mathbf{a}_r^{\mathsf{H}}(\theta_{m})\big)\biggr) \nonumber \\
 &=  \frac{2E_s}{\sigma^{2}}\tr\biggl(\mathbf{A}^{\mathsf{H}}\mathbf{A}\big(\mathbf{a}_t^{*}(\phi_{m})\mathbf{a}_t^{\mathsf{T}}(\phi_{m})\big) {\otimes}  \big(\mathbf{a}_r(\theta_{m})\mathbf{a}_r^{\mathsf{H}}(\theta_{m})\big)\biggr).
	 \end{align}
	
	This concludes the proof of the entries of $\mathbf{J}_{\boldsymbol{\theta}}$.
\end{IEEEproof}

\ifCLASSOPTIONcaptionsoff
  \newpage
\fi

\end{document}